\documentstyle[11pt]{article}

\title{Notes on highest weight modules of the \\
elliptic algebra ${\cal A}_{q,p}\left(\widehat{sl}_2\right)$}

\author{
Omar Foda\thanks{Department of Mathematics, University of Melbourne,
                         Parkville, Victoria 3052, Australia.},
Kenji Iohara\thanks{Department of Mathematics, Faculty of Science,
                            Kyoto University, Kyoto 606, Japan.},
Michio Jimbo$^{\dagger}$,
Rinat Kedem\thanks{Research Institute for Mathematical Sciences,
                            Kyoto University, Kyoto 606, Japan.},\cr
Tetsuji Miwa$^\ddagger$ and
Hong Yan\thanks{Institute of
Theoretical Physics, P. O. Box 2735, Beijing, China.} \cr
}

\date{}
\begin{document}
\maketitle
\begin{abstract}
We discuss a construction of highest weight modules for the recently
defined elliptic algebra ${\cal A}_{q,p}(\widehat{sl}_2)$, and make
several conjectures concerning them.  The modules are generated by the
action of the components of the operator $L$ on the highest weight vectors.
We introduce the vertex operators $\Phi$ and $\Psi^*$ through their
commutation relations with the $L$-operator. We present ordering rules
for the $L$- and $\Phi$-operators and find an upper bound for the
number of linearly independent vectors generated by them, which agrees
with the known characters of $\widehat{sl}_2$-modules.

\end{abstract}

\setcounter{section}{0}

%%%%%%%%%%%%%%%%%%%%%%%%%%%%%%%%%%%%%%%%%%%%%%%%%%%%%%%%%%
%\newfont{\g}{eufm9}

\renewcommand{\theequation}{\thesection.\arabic{equation}}
\newcommand{\gtg}{\mbox{\g g}}
\def\As{{\cal A}_{q,p}\bigl(\slth\bigr)}
\def\Ag{{\cal A}_{q,p}\bigl(\widehat{gl}_2\bigr)}
\def\e{\varepsilon}
\def\a(#1){{\cal A}_{q,p}({\widehat{#1}}_2)}
\def\nn{\nonumber}
\def\End{{\rm End}\,}
\def\Hom{{\rm Hom}\,}
\def\ch{{\rm ch}\,}
\def\itm#1{\par#1\quad}
\def\C{{\bf C}}
\def\be{\begin{equation}}
\def\en{\end{equation}}
\def\bea{\begin{eqnarray}}
\def\ena{\end{eqnarray}}
\def\beq{\begin{equation}}
\def\eeq{\end{equation}}
\def\beqa{\begin{eqnarray}}
\def\eeqa{\end{eqnarray}}
\def\refeq#1{(\ref{eqn:#1})}
\def\refto#1{\cite{#1}}
\def\lb#1{\label{eqn:#1}}
\def\rf#1{(\ref{eqn:#1})}
\def\disp{\displaystyle}
\def\Remark{\medskip\noindent {\sl Remark}\quad}
\def\Phim#1{\mathrel{\mathop{\kern0pt \Phi}\limits^#1}}
\def\Psim#1{\mathrel{\mathop{\kern0pt \Psi^*}\limits^#1}}
\def\Lm#1{\mathrel{\mathop{\kern0pt L^-}\limits^#1}}
\def\Lp#1{\mathrel{\mathop{\kern0pt L^+}\limits^#1}}
\def\Lpm#1{\mathrel{\mathop{\kern0pt L^\pm}\limits^#1}}

\def\C{{\bf C}}
\def\Z{{\bf Z}}
\def\H{{\cal H}}
\def\vep{\varepsilon}
\def\z{\zeta}
\def\mod{{\rm mod}~}
\def\qdet{\hbox{$q$-det}}
\def\id{{\rm id}}
\def\bra#1{\langle #1|}
\def\ket#1{|#1\rangle}
\def\Phim#1{\mathrel{\mathop{\kern0pt \Phi}\limits^#1}}
\def\Psim#1{\mathrel{\mathop{\kern0pt \Psi^*}\limits^#1}}
\def\LL#1{\mathrel{\mathop{\kern0pt L}\limits^#1}}
\def\Lm#1{\mathrel{\mathop{\kern0pt L^-}\limits^#1}}
\def\Lp#1{\mathrel{\mathop{\kern0pt L^+}\limits^#1}}
\def\Lpm#1{\mathrel{\mathop{\kern0pt L^\pm}\limits^#1}}
\def\glth{\widehat{gl}_2}
\def\slth{\widehat{sl}_2}
\def\A{{\cal A}}
\def\aqp{{\cal A}_{q,p}(\slth)}
\def\M{{\cal M}}
\def\MKL{{\cal M}_{k,l}}
\def\DMKL{{\cal M}^*_{k,l}}

\def\uq{U_q(\slth)}				% Uq(sl_2 hat)
\def\End{{\rm End}\,}
\def\Ad{{\rm Ad}}
\def\R{{\cal R}}

%\input xypic
%%%%%%%%%%%%%%%%%%%%%%%%%%%%%%%%%%%%%%%%%%%%%%%%%%%%%%%%%%

\setcounter{section}{0}
\setcounter{equation}{0}
\section{Introduction}
In our previous paper \cite{FIJKMY} we defined
the elliptic quantum affine algebra ${\cal A}_{q,p}({\hat g})$ with
$g=gl_2$ or $sl_2$.
The present paper is an attempt toward understanding the correct
elliptic analogues of the highest weight modules and vertex operators.
Our aim here is to
present conjectures concerning their existence and expected properties,
along with some experimental computations to support them.

\subsection{The elliptic algebra}
Let us recall the definition of the algebras.
The algebra
$\a(gl)$ is defined through a quadratic relation of the form
\beq
R^+(\z_1/\z_2)
\bigl(L(\z_1)\otimes{\rm id}\bigr)
\bigl({\rm id}\otimes L(\z_2)\bigr)
=\bigl({\rm id}\otimes L(\z_2)\bigr)
\bigl(L(\z_1)\otimes{\rm id}\bigr)
R^{*+}(\z_1/\z_2),\lb{DEF}\nn\\
\eeq
where $R^+(\z)=R^+(\z;p^{1/2},q^{1/2})$ is the elliptic $R$-matrix of
the eight vertex model, with an elliptic nome $p$ and a crossing
parameter $q$.  The matrix $R^{*+}$ is an elliptic $R$-matrix
with a scaled elliptic nome: $R^{*+}(\z)=R^+(\z;p^{*{1/2}},q^{1/2})$
with $p^{*{1/2}}=p^{1/2}q^{-c}$, where $q^{c/2}$ is a central element
of $\a(gl)$. The $L$-operator $L(\z)=\bigl(L_{\e\e'}(\z)\bigr)_{\e,\e'=\pm}$
is assumed to satisfy the parity relation $L_{\e\e'}(-\z)=\e\e'L_{\e\e'}(\z)$
and is expanded as
\beq\sum_\e L_{\e\e'}(\z)=\sum_{n\in\Z}L^{\e'}_n\z^{-n}.\eeq

The algebra $\a(sl)$ is defined by further requiring that the quantum
determinant be equal to $q^{c/2}$:
\begin{equation}
L_{++}(\z/q)L_{--}(\z)-L_{-+}(\z/q)L_{+-}(\z)=q^{c/2}.\lb{DET}
\end{equation}

\subsection{Highest weight modules}
A level-$k$ highest weight module $M$ of the algebra $\a(sl)$ is a
left $\a(sl)$-module with the action of the operators $L^\e_n$
satisfying the following properties:
\itm{(i)}
$M$ is graded, $M=\oplus_{d\ge0}M_d$, and $L^\e_n\in\Hom(M_d,M_{d-n})$.
\par\noindent
We say that a vector $w\in M$ has degree $d$ if $w\in M_d$.
\itm{(ii)}
Up to proportionality, there exists a unique vector of degree $0$,
$v\in M_0$, called the highest weight vector.
The whole module is generated by the highest weight vector: $M=\a(sl)v$.
\par\noindent
\itm{(iii)} The central element
$q^{c/2}$ acts as $q^{k/2}\times{\rm id}$ on $M$.
\par\noindent
The complex number $k$ is called the level of the representation.

These conditions are similar to those in the case
of the affine Lie algebras. For the latter,
there are two important classes of highest weight modules:

\itm{(a)}
Verma modules;

\noindent

\itm{(b)}
Irreducible highest weight modules.

\noindent
As is well known, these modules
have deformations to the case of quantized affine
algebras $U_q(\hat g)$,
the `trigonometric case', in such a way that the characters
remain the same.
We expect that the situation is analogous in the elliptic case.
We will introduce an elliptic analogue of the Verma module
(which we refer to by the same name), and give a spanning set of vectors.
If they are linearly independent, then the character is shown to be
unchanged from the case of Lie algebras.
The analogue of the level-one irreducible highest weight modules
has been discussed in \cite{FIJKMY}.
Assuming their existence, we give another conjectural basis for them,
using vertex operators.
Let us explain these constructions in more details.

As with the Lie algebra case,
the elliptic Verma modules carry two parameters,
the level-$k$ and a second parameter $l$ (see Section 3.1): $M={\cal M}_{k,l}$.
General vectors in ${\cal M}_{k,l}$ are created
by applying $L^\e_n$ repeatedly to the highest weight vector $v=|k,l\rangle$.
The set of vectors obtained this way contains redundancy.
In order to reduce the number of vectors we rewrite
the defining relation \refeq{DEF} and \refeq{DET} as normal-ordering rules
for the product of operators $L^\e_n$: Roughly speaking,
we bring $L^{\e_1}_{n_1}$
to the left of $L^{\e_2}_{n_2}$ if  $n_1<n_2$.
{}From this we find that the vectors
\begin{equation}\label{VEC}
w=L^{\e_1}_{n_1}\cdots L^{\e_m}_{n_m}|k,l\rangle
\qquad (n_1\leq\cdots\leq n_m<0, ~\e_j=\e_{j+1} \hbox{ if }n_j=n_{j+1}),
\end{equation}
constitute a spanning set.
The generating function of the numbers $N_d$ of the degree-$d$ vectors
in this set is equal to the character of the Verma modules for
$\widehat{sl}_2$:
\beq\label{Nd}
\sum_{d\geq0} N_d ~t^d = \prod_{n=1}^\infty{1\over(1-t^n)(1-t^{2n-1})}.
\eeq

In principle,
we can compute the action of $L^\e_n$ on $w$: Using the normal-ordering
rules, we can reduce
it to a linear combination of the vectors of the form
(\ref{VEC}).
However the normal-ordering rules involve infinite sums (see Section 3.2),
which makes the reduction process quite complicated.
We have not yet been able to show that the result is independent
of the order of reduction, which is necessary
to study the linear independence of (\ref{VEC}).

\subsection{Vertex operators}
A conjectural basis of the irreducible highest weight modules
(when the Verma module is reducible) is constructed differently.
We consider in this paper only the simplest case $k=1$.
As in the case of ${\widehat{sl}}_2$
we expect to obtain two level-one integrable irreducible
representations ${\cal H}^{(0)}$ and ${\cal H}^{(1)}$
as the quotients of the Verma modules ${\cal M}_{1,0}$ and
${\cal M}_{1,1}$, respectively. In the elliptic case, it is not
easy to describe these quotients explicitly in terms of the $L$-operators.
Instead we will use vertex operators to create the vectors in
${\cal H}^{(i)}$.

The type-I vertex operator $\Phi(\z)$ is defined similarly as in
\cite{FreResh}: We require that $\Phi$ intertwines $L$ and $R^+L$ (see
Section 4.1). This relation uniquely determines the action
of the vertex operator $\Phi^{(l\pm1,l)}_\e(\z)$:
${\cal M}_{k,l}\rightarrow{\cal M}_{k,l\pm1}$.
The vertex operator in the trigonometric case
\cite{FreResh} satisfies a quadratic commutation
relation (see also \cite{DFJMN} for the level-one case).
The determination of the coefficients in these relations
is based on an analysis of the $q$KZ equation,
of which an elliptic counterpart is yet unknown.
Nevertheless, we speculate that the elliptic vertex operator
satisfies a similar quadratic commutation relation.
In the level-one case, such a commutation relation is anticipated from the
identification of the vertex operator and the half-column transfer matrix
of the eight-vertex model (see \cite{8v}).

We will construct vectors in
${\cal H}^{(0)}\oplus{\cal H}^{(1)}$ by means of the vertex operators,
\[
\Phi^{(1-i,i)}_\e(\z):{\cal H}^{(i)}\rightarrow{\cal H}^{(1-i)}.
\]
Taking the parity of the vertex operator into account, we set
$$\sum_\e\Phi^{(1-i,i)}_\e(\z)=\sum_n\varphi^{(1-i.i)}_n\z^{-n}.$$
Starting from the highest weight vectors $|0\rangle$ and $|1\rangle$,
we create vectors of higher degrees by applying $\varphi^{(1-i,i)}_n$
repeatedly. We conjecture that all the vectors in ${\cal
H}^{(0)}\oplus{\cal H}^{(1)}$ are created in this way.  In order to
get a reduced set of vectors, we proceed as in the previous case: We
rewrite the commutation relation for the vertex operator as
normal-ordering rules. We also supplement the rules by a certain
inversion relation of $\Phi^{(1-i,i)}_\e(\z)$.  (This is analogous to
the $q$-determinant condition \refeq{DET}.)  If the above conjecture
is valid, the set of vectors
\[
\varphi^{(i,i+1)}_{n_1}\cdots\varphi^{(i+m-1,i+m)}_{n_m}|i+m\rangle
\]
where $n_1<\cdots<n_m<0$ and the indices $i$, $i+1$ , etc., are read
modulo $2$, span the space ${\cal H}^{(i)}$.

\vspace*{.2in}
The paper is organized as follows. In Section 2, we recall the
definition of the algebra $\aqp$. We define the Verma modules in
Section 3 and derive normal-ordering rules for the components of the
$L$-operator. We present a conjectural basis of the form (\ref{VEC})
and prove the equality (\ref{Nd}).  In Section 4, we define the vertex
operators in terms of their commutation relations with $L$, conjecture
a form for the commutation relations between $\Phi$ operators, and
define the type-II vertex operators $\Psi^*$ in terms of the operators
$L$ and $\Phi$ (for the physical interpretation of these operators,
see e.g. \cite{FIJKMY}).  In Section 5, we present ordering rules for
the components of the $L$ and $\Phi$, which determine the matrix
elements of the $\Phi$-components.  We write down normal-ordering
rules between components of the operator $\Phi$ and use this to
specify a different conjectural basis for the Verma module.  A
specialization to the case of level-one is considered in Section 6.
The text is supplemented by three appendices. For comparison and
motivation, we summarize in Appendix A the known results about the
$L$-operators and the vertex operators in the trigonometric case. In
Appendix B we give formulas for the two-point functions in the
elliptic case for various combinations of the $L$-operators and the
vertex operators.  In Appendix C, we discuss the form of the
Shapovalov-type determinants~\cite{KacKaz,DeCKac} which we
compute for small degree. These determinants have a simple factorized
form.

\setcounter{equation}{0}
\section{The algebra $\As$}

\subsection{$R$-matrix}
The purpose of this section is to recall the definition of the elliptic
algebra $\As$ \cite{FIJKMY}, thereby fixing the notation.

First we need to prepare the
$R$ matrix of the eight-vertex model given in \cite{Baxbk}.
In what follows we shall work with the parameters $p,q,\z$ which
are related to $I,I',\lambda,u$ in \cite{Baxbk}, equations.(10.4.23--24) by
\[
p=\exp(-\pi I'/I),\quad
q=-\exp(-\pi\lambda/2I),\quad
\zeta=\exp(\pi u/2I).
\]
The $R$-matrix has the structure
\begin{equation}\label{R}
R(\zeta)=
\left(
\begin{array}{cccc}
a(\zeta)&&&d(\zeta)\\
&b(\zeta)&c(\zeta)&\\
&c(\zeta)&b(\zeta)&\\
d(\zeta)&&&a(\zeta)
\end{array}
\right)
\end{equation}
with the entries given by the formulas
\begin{equation}\nonumber
\begin{array}{crc}
a(\z)=&\rho'& \Theta_{p^2}(pq^2) \Theta_{p^2}(q^2\z^2) \Theta_{p^2}(p\z^2),
\\
b(\z)=&\rho'q &\Theta_{p^2}(pq^2) \Theta_{p^2}(pq^2\z^2) \Theta_{p^2}(\z^2),
\\
c(\z)=&\rho'\z& \Theta_{p^2}(q^2) \Theta_{p^2}(pq^2\z^2) \Theta_{p^2}(p\z^2),
\\
d(\z)=&\rho'{\displaystyle \frac{p^{1/2}}{q\z}}&
\Theta_{p^2}(q^2) \Theta_{p^2}(q^2\z^2) \Theta_{p^2}(\z^2).
\end{array}
\end{equation}
Here we set
\[
\Theta_{p^2}(z)=(z;p^2)_\infty(p^2z^{-1};p^2)_\infty(p^2;p^2)_\infty
\]
and the infinite product symbol is defined as
\begin{eqnarray}
(z;p_1,\cdots,p_m)_\infty
&=&\prod_{n_1,\cdots,n_m\ge 0}(1-zp_1^{n_1}\cdots p_m^{n_m})\nonumber\\
&=&\exp\left(-\sum_{k=1}^\infty\frac{1}{(1-p_1^k)\cdots(1-p_m^k)}
\frac{z^k}{k}\right).\label{infpro}
\end{eqnarray}
The scalar $\rho'$
(related to $\rho$ in  \cite{Baxbk}, (10.4.24) via
$\rho'=-\rho p^{1/4}(q\z)^{-1}$) is a free parameter.
As in \cite{8v} we shall choose $\rho'$ as follows:
\begin{eqnarray}
&&\rho'=
\z^{-1}\frac{\xi(\z^2;p,q)}{\xi(\z^{-2};p,q)}
\frac{1}{(q^2\z^2;p)_\infty(pq^{-2}\z^{-2};p)_\infty
(p;p)^2_\infty(p^2;p^2)_\infty},
\nonumber\\
&&
\xi(z;p,q)
=\frac{(q^2 z;p,q^4)_\infty (pq^2 z;p,q^4)_\infty}
{(q^4 z;p,q^4)_\infty(p z;p,q^4)_\infty}.
\label{xi}
\end{eqnarray}
In the context of the eight-vertex model,
this choice
corresponds to normalizing the Boltzmann weights in such a way
that the partition function per site is $1$.

For the definition of the elliptic
algebra we introduce two more $R$ matrices
\begin{equation}\label{Rpm}
\begin{array}{ccl}
R^\pm(\z)&=\tau^\pm(\z)R(\z)&, \\
\tau^+(\z)&={\tau^-(\z^{-1})}^{-1}
&=
q^{-1/2}\z
{\displaystyle \frac{(q^2\z^2;q^4)_\infty}{(\z^2;q^4)_\infty}}
{\displaystyle \frac{(q^2\z^{-2};q^4)_\infty}{(q^4\z^{-2};q^4)_\infty}}.
\end{array}
\end{equation}
The corresponding entries will be denoted by $a^\pm(\z),b^\pm(\z)$, etc.
Later we will need to diagonalize them.
The eigenvalues are simply sums and differences of the entries, and
can be put into a factorized form
using the addition formula for theta functions.
We give below such formulas,
noting that $a^\pm(\z),b^\pm(\z)$ are even
and $c^\pm(\z),d^\pm(\z)$ are odd in $\z$.
\begin{equation}
\label{R-entries}
\begin{array}{rcl}
a^{\pm}(\zeta) + d^{\pm}(\z) &=& \rho(\z^{\pm2})^{\pm 1}
\displaystyle
\frac{\overline{\alpha}(\z^{-1})}{\overline{\alpha}(\z)}~,\\
b^{\pm}(\z) + c^{\pm}(\z) &=& \rho(\z^{\pm2})^{\pm1} q^{\pm 1}
\displaystyle
\frac{1 + ( q^{-1} \z )^{\pm 1}}{1 + (q \z)^{\pm 1}}
\displaystyle
\frac{\overline{\beta}(\z^{-1})}{\overline{\beta}(\z)}~.
\end{array}
\end{equation}
Here
\begin{eqnarray}
\rho(z)&=&q^{-1/2}\displaystyle
\frac{(q^2z;q^4)_\infty^2}
{(z;q^4)_\infty(q^4z;q^4)_\infty},
\label{rho}\\
\overline{\alpha}(\zeta)&=&\displaystyle
\frac{(p^{1/2}q\zeta; p)_\infty}
{(p^{1/2}q^{-1}\zeta; p)_\infty}
\displaystyle \frac{(p q^4\zeta^2;p,q^4)_\infty(p\zeta^2;p,q^4)_\infty}
{(pq^2\zeta^2;p,q^4)^2_\infty},
\label{alpha}\\
\displaystyle \overline{\beta}(\zeta)&=&\displaystyle
\frac{(-pq\zeta;p)_\infty}{(-pq^{-1}\zeta;p)_\infty}
\displaystyle \frac{(pq^4\zeta^2;p,q^4)_\infty(p\zeta^2;p,q^4)_\infty}{
(pq^2\zeta^2;p,q^4)^2_\infty}.
\label{beta}
\end{eqnarray}

{}From now on we regard $p^{1/2},\z$ as indeterminates.
In \cite{FIJKMY}  $q^{1/2}$ was also treated as a formal variable.
In this paper we take $q^{1/2}$ to be a fixed complex number such that
$0<|q^{1/2}|< 1$.
We define $R^\pm(\z)$ by (\ref{R-entries}) to be a formal series
\[
R^\pm(\z)=\sum_{n\in \Z} R^\pm_n \z^n
\]
whose coefficients are
in the ring $B=\C[[p^{1/2}]]$ of formal power series in $p^{1/2}$,
wherein we specify the factor $(1 + (q \z)^{\pm 1})^{-1}$ in
$b^\pm(\z)+c^\pm(\z)$ to be an expansion in powers of $\z^{\pm 1}$.
The Taylor coefficients of
$\overline{\alpha}(\z)=\sum_{j\ge 0} \alpha_j\z^j$
satisfy $\alpha_j\in p^{j/2}B$, which ensures that the product
$\overline{\alpha}(\z^{-1})/\overline{\alpha}(\z)$ is a  well defined formal
series with coefficients in $B$.
Likewise for $\overline{\beta}(\z)$.
The matrices $R^\pm$ therefore satisfy
\begin{equation}\label{estiR}
R^\pm_n\equiv 0~~ \mod  \bigl(p^{1/2}\bigr)^{\max\,(\mp n,0)}\!B
\qquad \forall n\in\Z.
\end{equation}
In particular, when $p=0$,
$R^+(\z)$ (resp. $R^-(\z)$) contains only non-negative
(resp. non-positive) powers in $\z$.

We regard $R^\pm(\z)$ as linear operators on $V\otimes V$, with
$V=Bv_+\oplus Bv_-$, and set $R^\pm(\z)v_{\vep'_1}\otimes v_{\vep'_2}
=\sum v_{\vep_1}\otimes v_{\vep_2}
R^\pm(\z)_{\vep_1\vep_2;\vep_1'\vep_2'}$.
When written in the matrix form as in (\ref{R}),
the entries of $R^\pm$ are arranged in the order
$(\vep_1,\vep_2)=(++),(+-),(-+),(--)$.

We list below the main properties for
the $R^\pm$-matrices which are relevant in the sequel.

\begin{description}
\item[Yang-Baxter equation]
\begin{equation}\label{YBE}
R^\pm_{12}(\z_1/\z_2)R^\pm_{13}(\z_1/\z_3)R^\pm_{23}(\z_2/\z_3)
=
R^\pm_{23}(\z_2/\z_3)R^\pm_{13}(\z_1/\z_3)R^\pm_{12}(\z_1/\z_2),
\end{equation}
\item[Unitarity]
\begin{equation}\label{unitarity}
R^\pm_{12}(\z_1/\z_2)R^\mp_{21}(\z_2/\z_1)=\id,
\end{equation}
\item[Crossing symmetry]
\begin{equation}\label{crossing}
R^\pm_{21}(\z_2/\z_1)^{t_1}=
\sigma^x_1 R^\mp_{12}(-q^{-1}\z_1/\z_2)\sigma^x_1,
\end{equation}
\item[Quasi-periodicity]
\begin{equation}\label{pro1}
R^\pm_{12}(-\zeta)=\sigma_1^zR^\pm_{12}(\zeta)\sigma_1^z=
\sigma^z_2R^\pm_{12}(\zeta)\sigma^z_2~,
\end{equation}
\begin{equation}\label{pro2}
R^+_{12}(-p^{1/2}\zeta)=\sigma_1^xR^-_{12}(\zeta)\sigma_1^x=
\sigma^x_2R^-_{12}(\zeta)\sigma^x_2~.
\end{equation}
\end{description}
Here, if $R^\pm(\z)=\sum a_i\otimes b_i$, with $a_i,~b_i
\in {\rm End}(V)$, then
$R^\pm_{13}(\z)=\sum a_i\otimes \id \otimes b_i$,
$R^\pm_{21}(\z)^{t_1}=\sum b_i^t\otimes a_i$,
etc., where
$b_i^t$ signifies the transpose of $b_i$.
The Pauli matrices are chosen to be
\[
\sigma^x=\left(\matrix{0&1\cr 1&0\cr}\right),
\qquad
\sigma^y=\left(\matrix{0&-i\cr i&0\cr}\right),
\qquad
\sigma^z=\left(\matrix{1&0\cr 0&-1\cr}\right),
\]
the suffix $j(=1,2)$
indicating that they are acting on the $j^{\rm th}$-component.

\subsection{The elliptic algebra}
The elliptic algebra is defined in terms of generators and relations.
Consider the formal series of the form
\begin{equation}
L(\zeta)=\left(
\begin{array}{cc}
L_{++}(\zeta)&L_{+-}(\zeta)\\
L_{-+}(\zeta)&L_{--}(\zeta)
\end{array}\right)
\end{equation}
where
\begin{equation}\label{L-entries}
L_{\vep\vep'}(\zeta)=\sum_{n\in\Z} L_{\vep\vep',n}\z^{-n},
\qquad
L_{\vep\vep',n}=(-p^{1/2})^{\max\,(n,0)}\overline{L}_{\vep\vep',n}.
\end{equation}
Here
$\overline{L}_{\vep\vep',n}$ ($n\in$ {\bf Z}, $\vep,\vep'=\pm$,
$\vep\vep'=(-1)^n$)
are abstract symbols.
By convention we set
\begin{equation}\label{ParL}
\overline{L}_{\vep\vep',n}=0
\qquad \hbox{ if }\vep\vep'\neq (-1)^n.
\end{equation}
In later sections we will use $L^{\vep'}_n$ to represent the homogeneous
components of $L(\z)$:
\begin{equation}\nonumber
L_{+\vep'}(\z)+L_{-\vep'}(\z)=\sum_{n\in\Z}L^{\vep'}_n\z^{-n}.
\end{equation}

We define ${\cal A}_{q,p}(\glth)$ to be the
topological $B$-algebra
with generators $\overline{L}_{\vep\vep',n}$
and an invertible central element $q^{c/2}$, through the
following defining relations:
\begin{equation}\label{RLL}
R^+_{12}(\zeta_1/\zeta_2)\LL{1}(\zeta_1)\LL{2}(\zeta_2)
=\LL{2}(\zeta_2)\LL{1}(\zeta_1)R^{*+}_{12}(\zeta_1/\zeta_2)~,
\end{equation}
where
\[
\LL{1}(\zeta)= L(\z)\otimes  {\rm id},\qquad
\LL{2}(\zeta)=  {\rm id} \otimes L(\z),
\]
and
\begin{equation}
R^{*+}(\z)=R^+(\z;p^{*1/2},q^{1/2}),
\qquad p^{*1/2}=p^{1/2}q^{-c}.
\end{equation}
To be precise, let ${\cal U}'$ denote the tensor algebra over $B$
on the letters $\overline{L}_{\vep\vep',n}$ with the central element
$q^{c/2}$ adjoined,
${\cal U}$ its $p^{1/2}$-adic completion,
and
${\cal I}$ the ideal generated by the coefficients (in $\z_1,\z_2$)
of the matrix entries of the LHS$-$RHS of (\ref{RLL}).
Then $\Ag={\cal U}/\overline{\cal I}$,
where $\overline{\cal I}$ signifies the closure of ${\cal I}$.

We wish to verify here that the both sides of (\ref{RLL}) are
indeed well defined.
The left hand side has the form
\[
\sum_{m,n\in\Z}L(m,n)\z_1^{-m}\z_2^{-n},
\quad
L(m,n)=\sum_{j\in\Z}R_j\left(L_{m+j}\otimes 1\right)
\left(1\otimes L_{n-j}\right).
\]
Thanks to (\ref{estiR}) and (\ref{L-entries}),
$L(m,n)$ is convergent in the $p^{1/2}$-adic
topology and moreover we have
\[
L(m,n)\equiv 0~~ \mod  \bigl(p^{1/2}\bigr)^{\max\,(m,0)}\!B
\qquad \forall m,\,n\in\Z.
\]
Similarly the right hand side has the form
\[
\sum_{m,n\in\Z}R(m,n)\z_1^{-m}\z_2^{-n},
\qquad
R(m,n)=\sum_{j\in\Z}
\left(1\otimes L_{n-j}\right)\left(L_{m+j}\otimes 1\right)R_j
\]
with
\[
R(m,n)
\equiv 0~~ \mod  \bigl(p^{1/2}\bigr)^{\max\,(m,0)}\!B.
\]

It follows from this estimate that
(\ref{RLL}) multiplied by a common scalar $\sum_{j\in\Z}c_j(\z_1/\z_2)^j$
is valid as a formal series relation provided $c_j\in p^{\max(0,-j)}B$.
In particular the common scalar $\rho(\z_1^2/\z_2^2)$ can be divided out from
(\ref{RLL}).
This remark will be used when we discuss the normal-ordering rules in
Section 3.

Let $V_\xi=V\otimes B[\xi,\xi^{-1}]$, and
define $\pi_\xi:\Ag\rightarrow \End(V_\xi)$ by
\[
\pi_\xi\left(L_{\vep_1\vep_1'}(\z)\right)v_{\vep_2'}=
\sum_{\vep_2} v_{\vep_2}R^+(\z/\xi)_{\vep_1\vep_2;\vep_1'\vep_2'},
\qquad \pi_\xi\left(q^{c/2}\right)=1.
\]
The Yang-Baxter equation for $R^+$ ensures that $V_\xi$ is an $\Ag$-module.
This is analogous to the evaluation module of the quantum affine algebras.

\subsection{The quantum determinant}
The algebra $\As$ is defined by imposing further a relation on the
quantum determinant to be defined below.
First note that the $R$ matrix (\ref{R}) at $\z=-q^{-1}$ has the simple form
\begin{equation}
R(-q^{-1})=
\left(\matrix{0 & 0 & 0 & 0 \cr
              0 & 1 & 1 & 0 \cr
              0 & 1 & 1 & 0 \cr
              0 & 0 & 0 & 0 \cr
}\right).\label{SPVL}
\end{equation}
Thus $R(-q^{-1})$ has rank one, and its image is spanned by
$w=v_+\otimes v_- + v_-\otimes v_+$.
Applying both sides of (\ref{RLL}) to $w$ and specializing to
$\z_1=-q^{-1}\z_2$, we see that $\LL{2}(\z)\!\LL{1}(-q^{-1}\z)w$
lies in the image of $R_{12}(-q^{-1})$.
Therefore it can be written as
\begin{equation}\label{LLw}
\LL{2}(\z)\!\LL{1}(-q^{-1}\z)\,w= \qdet L(\z)\, w
\end{equation}
where $\qdet L(\z)$, called the quantum determinant,
is  a certain series with coefficients in the algebra $\Ag$.
Writing down the relation (\ref{RLL}) at
$\z_1=-q^{-1}\z_2$ we find the
following equivalent expressions for $\qdet L(\z)$:
\begin{equation}\label{qdet}
\begin{array}{ccccc}
\qdet L(\z)&=& L_{++}(\z/q)L_{--}(\z)&-&L_{-+}(\z/q)L_{+-}(\z) \cr
&=& L_{--}(\z/q)L_{++}(\z)&-&L_{+-}(\z/q)L_{-+}(\z) \cr
&=& L_{++}(\z)L_{--}(\z/q)&-&L_{+-}(\z)L_{-+}(\z/q) \cr
&=& L_{--}(\z)L_{++}(\z/q)&-&L_{-+}(\z)L_{+-}(\z/q) .\cr
\end{array}
\end{equation}
We have in addition
\begin{eqnarray}
L_{++}(\z/q)L_{-+}(\z)&=&L_{-+}(\z/q)L_{++}(\z),
\nonumber\\
L_{++}(\z)L_{+-}(\z/q)&=&L_{+-}(\z)L_{++}(\z/q),
\nonumber\\
L_{+-}(\z/q)L_{--}(\z)&=&L_{--}(\z/q)L_{+-}(\z),
\nonumber\\
L_{--}(\z)L_{-+}(\z/q)&=&L_{-+}(\z)L_{--}(\z/q).
\nonumber
\end{eqnarray}
Despite the appearance of two different $R$ matrices in the defining relation
(\ref{RLL}), the usual argument applies to
show that all the coefficients of $\qdet L(\z)$ belongs to the center of $\Ag$.
Let us verify this statement.
Using (\ref{RLL}) repeatedly and specializing the parameter, we have
for any $v\in V$
\begin{eqnarray}
&&\LL{3}(\z')\LL{2}(\z)\LL{1}(-q^{-1}\z)\,
R^{+*}_{23}(\z/\z')R^{+*}_{13}(-q^{-1}\z/\z')\, w\otimes v
\nonumber\\
=&&R^{+}_{23}(\z/\z')R^{+}_{13}(-q^{-1}\z/\z')\,
\LL{2}(\z)\LL{1}(-q^{-1}\z) \LL{3}(\z')\, w\otimes v.
\label{LLL}
\end{eqnarray}
On the other hand,
using the unitarity (\ref{unitarity}) and crossing symmetry (\ref{crossing})
one can verify that $\qdet L(\z)$ acts as $1$ on the module $V_\xi$.
Taking the image of (\ref{LLw}) on $V_{\z'}$ we have for any $v\in V_{\z'}$
\[
R^+_{23}(\z/\z')R^+_{13}(-q^{-1}\z/\z')\, w\otimes v =  w\otimes v.
\]
Along with (\ref{LLw}) we conclude from (\ref{LLL}) that
\[
\LL{3}(\z')\,\qdet L(\z)\, w\otimes v =
\qdet L(\z)\!\LL{3}(\z')\, w\otimes v,
\]
which means $\qdet L(\z)$ commutes with all the entries of $L(\z')$.

Imposing further the relation
$\qdet L(\z)= q^{c/2}$ we define the quotient algebra
\[
\aqp={\cal A}_{q,p}(\glth)/\langle \qdet L(\z)- q^{c/2}\rangle.
\]
We note that $V_\xi$ is also a module over $\As$ since
$\qdet L(\z)$ acts as $1=q^{c/2}$.
In general we say that an $\As$-module has level $k\in \C$
if the central element $q^{c/2}$ acts as a scalar $q^{k/2}$.
Thus $V_\xi$ has level $0$.

Unlike in the trigonometric case one cannot define the notion of
weights since the $R$-matrix does not have a spin-conservation
property, i.e.
does not commute with matrices of the form $h\otimes h$ with
$h$ an arbitrary diagonal matrix.
Nevertheless ${\cal A}_{q,p}(\glth)$ can be made a
$\Z$-graded topological algebra by setting
\[
\deg L_{\vep\vep',n}=-n, \qquad \deg q^{c/2}=0,
\]
which corresponds to the principal grading for
affine Lie algebras.
Clearly $\aqp$ inherits this grading as well.

\medskip\noindent
{\sl Remark.\quad} The degeneration of the elliptic algebra in the
trigonometric limit $p\rightarrow 0$ has been discussed in \cite{FIJKMY}.
In the present normalization, the elliptic $L$-operator is
formally related to the trigonometric $L^\pm$-operators in Appendix A via
\begin{eqnarray*}
L^+(\z)&=&q^{-c/4}L(q^{c/2}\z)\Bigl|_{p=0},
\\
L^-(\z)&=&q^{-c/4}\sigma^xL(-p^{1/2}\z)\sigma^x\Bigl|_{p=0}.
\end{eqnarray*}

\setcounter{section}{2}
\setcounter{equation}{0}

\section{Verma modules and their spanning vectors}

\subsection{Definition of the Verma module}

We expect that the infinite dimensional representations of $\aqp$ are
smooth deformations of the corresponding representations which exist
in the trigonometric limit.  In the case of the quantized affine
algebra $U_q(\widehat{sl}_2)$, there exist level-$k$ Verma modules
$\M_{k,l}^0$, with highest weights $(k - l) \Lambda_0 + l \Lambda_1$,
where $k$ and $l$ are arbitrary complex numbers. We wish to define the
Verma modules $\M_{k,l}$ over $\aqp$ in an analogous way. It is the
universal module of level $k$ having a cyclic vector $\ket{k,l}$, such
that $\M_{k,l}=\aqp|k,l\rangle$, with the properties
\begin{eqnarray}
&&L^\e_{n } |k, l\rangle = 0~,~~\forall n>0,
\nonumber\\
&& L^+_{0} |k, l\rangle = a q^{(k-l)/2} |k, l\rangle~,\qquad
{}~L^-_{0} |k, l\rangle = a q^{   l /2} |k, l\rangle.
\label{hwcond}
\end{eqnarray}
Here the scalar $a$ accommodates the condition that the quantum
determinant acts as $q^{k/2}$. {}From the results for the two-point
functions (see Appendix B) we find that one should choose
\cite{FIJKMY}
\be\label{a-const}
a^2=\frac{\overline{\beta}(-q)}{\overline{\beta}^*(-q)},
\en
where $\overline{\beta}(\z)$ is defined in (\ref{R-entries}) and
$\overline{\beta}^*(\z)$ is the same function with $p^{1/2}$ replaced
by $p^{*1/2}=p^{1/2}q^{-k}$.  The normalization (\ref{hwcond}) is
discussed further in Appendix B.

The precise definition of $\M_{k,l}$ is as follows.  Let ${\cal J}$
denote the left ideal of $\aqp$ generated by all $L^\e_{n}$ with
$n>0$, together with $L^+_{0}-aq^{(k-l)/2}$, $L^-_{0}-aq^{l/2}$.  Then
we define $\M_{k,l}=\aqp/\bar{\cal J}$ and
$\ket{k,l}=1~\bmod~\bar{\cal J}$, where $\bar{\cal J}$ signifies the
$p^{1/2}$-adic closure.

Clearly, $\M_{k,l}$ admits a $\Z$-grading coming from that of $\aqp$,
$\M_{k,l}=\oplus_{d\in\Z}\bigl(\M_{k,l}\bigr)_d$, where $\oplus$ is a
topological direct sum. Explicitly, the grading is given by the degree
\be
d={\rm deg}    \Bigl( L^{\e_1}_{n_1}\cdots L^{\e_m}_{n_m} |k, l\rangle \Bigr)
= -  \sum_{j=1}^m n_j.
\en

In Section 3.2 we will show that the set of degree-$d$ vectors
\be\label{LBASIS}
       L^{\e_1}_{n_1} \cdots L^{\e_m}_{n_m} | k, l \rangle
\en
where $m\ge 0$,
\be\label{LBcond1}
n_1\le\cdots\le n_m<0
\en
and
\be\label{LBcond2}
\e_j=\e_{j+1}~\hbox{ if }~n_j=n_{j+1},
\en
is a spanning set of $\left(\M_{k,l}\right)_d$.
This provides us with an upper bound for the character of $\MKL$,
\be\label{M-char}
{\rm ch}\,\M_{k, l}~{\buildrel{\rm def}\over=}~
\sum_{d\in\Z}{\rm dim}\,\bigl(\M_{k,l}\bigr)_d~t^d.
\en

We conjecture that the vectors (\ref{LBASIS})
with the restrictions (\ref{LBcond1}) and (\ref{LBcond2})
are in fact all linearly independent.
As we show in Section 3.3, this is equivalent to stating
that the character (\ref{M-char}) is given by the same expression as the
(principal) character for the Verma module of the affine Lie
algebra $\slth$:
\be\label{v-char}
{\rm ch}\,\M_{k, l}
=\prod_{n=1}^\infty{\displaystyle1\over\displaystyle (1-t^n)(1-t^{2n-1})}.
\en

In order to show that (\ref{LBASIS}) span $\M_{k,l}$, we will derive
in the next Section normal-ordering rules for the Laurent
components of $L$, which can be used to reduce an arbitrary vector in
$\bigl(\M_{k, l}\bigr)_d$ to a linear combination of vectors of the
form (\ref{LBASIS}) with the restrictions (\ref{LBcond1}) and
(\ref{LBcond2}).

One can define the right Verma module $\M_{k,l}^*=\bra{k,l}\aqp$
in the same way, requiring that
\begin{eqnarray*}
&&\bra{k,l}L^{\e}_{n } = 0~,~~\forall n<0,
\nonumber\\
&&\bra{k,l} L^+_{0} = \bra{k,l}a q^{(k-l)/2}~,\qquad
{}~\bra{k,l}L^-_{0} = \bra{k,l}a q^{l /2}.
\end{eqnarray*}
The normal-ordering rules allow us to reduce the pairing of
$\bra{k,l}u$ and $v\ket{k,l}$ ($u,v$ being monomials in $L^\e_{n}$)
to a scalar.
We expect (without proof)
that the result is independent of the order of the reduction,
and that the resulting bilinear map
\[
\M^*_{k,l}\times \M_{k,l}\rightarrow B
\]
gives a non-degenerate pairing for generic $k,l$.
As usual,  we denote the pairing of $\bra{k,l}u$ and
$v\ket{k,l}$, $u,v\in \aqp$ by
$\langle uv \rangle=\bra{k,l}uv\ket{k,l}$, and refer to it
as the expectation value of $uv$.

\subsection{Normal-ordering of $L$-operators}
It is possible to rewrite the defining relation (\ref{RLL}) of
$\aqp$ in the form of normal-ordering rules, i.e.
for any  $n_1>n_2$ there is an equality of the form
\[\label{LLnormal}
L^{\e_1}_{n_1}L^{\e_2}_{n_2}
=\sum_{\sigma_1,m_1,\sigma_2,m_2\atop
m_1+m_2=n_1+n_2,m_1\le m_2}
C^{\e_1,n_1,\e_2,n_2}
_{\sigma_1,m_1,\sigma_2,m_2}
L^{\sigma_1}_{m_1}L^{\sigma_2}_{m_2}.
\]
To show this, we use a basic idea presented in
\cite{LW,LP}, which we will now outline in a simpler context
discussed there.

Consider the following commutation relation for
an operator $z(\z)=\sum_{n\in\Z}z_n\z^{-n}$:
\be\label{LWEQ}
\left(1-{\z_2\over\z_1}\right)^{-2/k}
z(\z_1) z(\z_2)
=\left(1-{\z_1\over\z_2}\right)^{-2/k}
z(\z_2) z(\z_1),
\en
where $k$ is an arbitrary complex number.
Define the coefficients $a_n$ and ${\widetilde a}_n$ by
\[
(1-\z)^{-2/k}=\sum_{n\ge0}a_n\z^n={\displaystyle1\over\displaystyle
\sum_{n\ge0}{\widetilde a}_n\z^n},
\]
and let
\[
Z(n_1,n_2)=\sum_{i\ge0}a_iz_{n_1-i}z_{n_2+i}~.
\]
This is a linear combination of $z_{m_1}z_{m_2}$ such that
$m_1+m_2=n_1+n_2$ and $m_1\le n_1$. Conversely, $z_{m_1}z_{m_2}$
is written as a linear combination of $Z(n_1,n_2)$ such that
$n_1+n_2=m_1+m_2$ and $n_1\le m_1$:
\[
z_{m_1}z_{m_2}=\sum_{i\ge0}{\widetilde a}_iZ(m_1-i,m_2+i).
\]
Equation (\ref{LWEQ}) implies that
\be\label{LWNO}
Z(m_1,m_2)=Z(m_2,m_1).
\en
rewriting $z_{n_1}z_{n_2}$ as a linear combination of
$Z(m_1,m_2)$, using (\ref{LWNO}) for the terms for which $m_1>m_2$,
$z_{n_1}z_{n_2}$ can be expressed as a linear combination of
$z_{m_1}z_{m_2}$ such that $m_1+m_2=n_1+n_2,m_1\le m_2$.

In the above argument
infinite sums of the operators $z_{m_1}z_{m_2}$ appear.
In \cite{LW}, and also in the present context discussed below,
these are well-defined as operators acting on highest
weight modules.

To obtain the normal-ordering rules for the $L$-operators from (\ref{RLL}),
we proceed as above, with the main differences being the following:

1. The defining relation (\ref{RLL}) is a matrix equation;
we reduce it to scalar equations by diagonalization of the $R$-matrix.

2. The coefficients of these scalar equations are
Laurent series rather than power series in $\z=\z_1/\z_2$ or
$\z^{-1}$. We must therefore
factorize the dependence on positive and negative
powers of $\z$ (the Riemann-Hilbert splitting),
and redistribute the factors
appropriately so that we obtain a formal power series in
$\z$ on one side, and $\z^{-1}$ on the other side of each
equation, just as in the simpler example above.

In fact, in (\ref{R-entries})
we have already prepared the diagonalization and the factorization.
As noted in Section 2.2,
the common factor $\rho(\z^2)$ can be factored out of (\ref{RLL}),
which is then
rewritten as
\be
{h^{(1)\sigma}(\z_2/\z_1)\over h^{*(1)\tau}(\e\z_2/\z_1)}
L_{\sigma,\tau,\e}(\z_1,\z_2)
=c(\e,\tau){h^{(2)\sigma}(\z_1/\z_2)\over h^{*(2)\tau}(\e\z_1/\z_2)}
L_{\sigma,\tau,\e}(\z_2,\z_1),\label{RLLf}
\en
where
$$
L_{\sigma,\tau,\e}(\z_1,\z_2)=
\sum_{\e',\e''}c(\e,\e'')L_{\e'\e''}(\z_1)
L_{\sigma\e'\,\tau\e''}(\z_2),$$
\beq\label{h-def}
h^{(1)+}(\z)=h^{(2)+}(\z)=\overline{\alpha}(\zeta),~~
h^{(1)-}(\z)=\overline{\beta}(\z),~~h^{(2)-}(\z)={1+q\z\over
q+\z}\overline{\beta}(\z),
\eeq
and
$$c(\e,\e')=\cases{-1&if $\e=\e'=-;$\cr+1&otherwise.\cr}$$
Again, the symbol $*$ is used to indicate the substitution
$p^{1/2}\rightarrow p^{*1/2}$.

Define the coefficients $h^{(i)\sigma,\tau,\e}_n$
and ${\widetilde h}^{(i)\sigma,\tau,\e}_n$ as
\[
{h^{(i)\sigma}(\z)\over h^{*(i)\tau}(\e\z)}=
\sum_{n\ge0}h^{(i)\sigma,\tau,\e}_n\z^n=
{\displaystyle1\over\displaystyle\sum_{n\ge0}
{\widetilde h}^{(i)\sigma,\tau,\e}_n\z^n}.
\]
Then, from (\ref{RLLf}) we have, for $m>n$,
\begin{eqnarray}
&&\sum_\e c(\e,\e')~L^\e_mL^{\tau\e}_n
=\sum_{i\ge0\atop\e}c(\e,\e')~
L^\e_{\left[{m+n\over2}\right]_--i}
L^{\tau\e}_{\left[{m+n\over2}\right]_++i}
\nonumber\\
&&~\times\Biggl\{
\sum_{j<0}h^{(1)\sigma,\tau,\e'}_{i+\delta+j}~
{\widetilde h}^{(1)\sigma,\tau,\e'}_{\left[{m-n\over2}\right]_--j}
+c(\tau,\e')\sum_{j\ge0}h^{(2)\sigma,\tau,\e'}_{i-j}~
{\widetilde h}^{(1)\sigma,\tau,\e'}_{\left[{m-n\over2}\right]_--j}\Biggr\},
\nonumber\\
\label{LLnormal-order}
\end{eqnarray}
where
\begin{eqnarray*}
&&\delta=\cases{0&if $m\equiv n~(2)$;\cr1&if $m\not\equiv n~(2)$,}\qquad
\sigma=\cases{\tau&if $m\equiv n~(2)$;\cr-\tau&if $m\not\equiv n~(2)$,}
\end{eqnarray*}
and
\[
\left[\frac{m}{2}\right]_\pm
=\cases{m/2&if $m$ is even;\cr(m\pm1)/2&if $m$ is odd.}
\]

Using the relation (\ref{LLnormal-order}), we obtain from the quantum
determinant condition (\ref{qdet}) the additional relation
\begin{eqnarray}
&&\sum_\e c(\e,\e')~L_n^{\e}L_n^{-\e}~
\Bigl(1+\e'\Bigl(\frac{\overline{\beta}^*(-\e'
q)}{\overline{\beta}(-q)}-1\Bigr)
\Bigr)=2\delta_{\e',+}\delta_{n,0}q^{k/2}\nonumber\\
&&\qquad -\sum_{{m>0}\atop{\e=\pm}}c(\e,\e')~L^{\e}_{n-m}L^{-\e}_{n+m}
{}~\Bigl((-q)^{-m}+\sum_{i>0}(-q)^i d_{i,m}^{\e,\e'}\Bigr)~,\label{qdetnorm}
\end{eqnarray}
where
\[
d_{m,i}^{\e,\e'}=\sum_{j<0}h^{(1)-,-,\e'}_{m+j}
\widetilde{h}^{(1)-,-,\e'}_{i-j}+
\e'\sum_{j\geq0}h^{(2)-,-,\e'}_{m-j}\widetilde{h}^{(1)-,-,\e'}_{i-j}~.
\]
The relations (\ref{LLnormal-order}) and (\ref{qdetnorm}) imply that
products of the form $L_m^\e L_n^{\e'}$ with $m>n$, or with $m=n$ and
$\e=-\e'$, can be expressed as sums of products with $m\leq n$, with
$\e=\e'$ if $m=n$.  Therefore we see that the module $\MKL$ is spanned
by vectors of the form (\ref{LBASIS}) with $n_j$ satisfying
(\ref{LBcond1}) and $\e'_j$ satisfying (\ref{LBcond2}).
The module $\DMKL$ is similarly spanned by
\be\label{DLbasis}
\langle k,l| L_{-n_m}^{\e_m}\ldots L_{-n_2}^{\e_2} L_{-n_1}^{\e_1},
\en
with $n_j$ and $\e_j$ again as in (\ref{LBcond1}) and (\ref{LBcond2}).

We note that the determinant $\Delta_d$ of the matrix formed by
coupling all degree-$d$ vectors of the form (\ref{DLbasis}) with
(\ref{LBASIS}) can be computed for small $d$, and has a simple
factorized form reminiscent of the Shapovalov determinant
\cite{KacKaz,DeCKac}. We discuss this factorization in Appendix C.

\subsection{Counting states}
We now show that the number of vectors of the form
(\ref{LBASIS})--({\ref{LBcond2})
is equal to the number of linearly independent vectors of fixed degree
in the Verma module of $\slth$.
This places the upper bound (\ref{v-char}) on the character of $\MKL$.

We call
\[
S=L^{\e_1}_{n_1}\cdots L^{\e_n}_{n_m}
\]
a string if it satisfies $n_1\le\cdots\le n_m<0$.
It is a degree-$d$ string if $-\sum_{j=1}^m n_j=d$.
We wish to show that the generating function of the numbers of
the degree-$d$ strings which satisfy the adjacency conditions

(i) if $n_j=n_{j+1}$ then $\e_j=\e_{j+1}$,

\noindent
is equal to (\ref{v-char}).

Let us first introduce three terms:
An A-string is a string $S$ satisfying (i), and
a B-string is a string $S$ satisfying
the following two conditions:

(ii) if $\e_j=+$ then $n_j$ is odd;

(iii) if $n_j=n_{j+1}$ then $(\e_j,\e_{j+1})\ne(-,+)$.

\noindent Finally a C-string is a string $S$ satisfying (iii).
Note that the A-strings are  C-strings which satisfy (i), and
the B-strings are  C-strings which satisfy (ii).
Note also that given a set of indices $\{\e_j,n_j\}_{j=1}^m$,
where $\e_j=\pm$ and $n_j<0$,
there is a unique way of relabeling them to get a C-string.

It is straightforward to see that the generating function of the number of
degree-$d$ B-strings is equal to the  character (\ref{v-char}). We will
show that the number of the degree-$d$ A-strings is equal to
the number of the degree-$d$ B-strings by constructing
successions of degree-$d$ C-strings that connect
A-strings with B-strings in one to one correspondence.

Let us introduce two moves on the C-strings:
\par\noindent
Given a C-string $S$ that is not an A-string,
an A-move is to modify $S$ to $S'$ by the following procedure.

Scan the string $S$ from the right for $j$ such that $\e_j\ne\e_{j+1}$
and $n_j=n_{j+1}$. Then, we have necessarily that $\e_j=+$ and $\e_{j+1}=-$.
Remove $L^{+}_{n_j}$ and $L^{-}_{n_{j+1}}$
from $S$ and add $L^+_{2n_j}$ to get a new C-string. This is $S'$.
\par\noindent
Given a C-string $S$ that is not a B-string,
an B-move is to modify $S$ to $S''$ by the following procedure.

Scan the string $S$ from the left for $j$ such that $\e_j=+$ and
$n_j$ is even. Remove $L^{+}_{n_j}$
from $S$ and add $L^+_{n_j/2}$  and $L^-_{n_j/2}$
to get a new C-string. This is $S''$.

Note that $S\rightarrow S'$ is an A-move if and only if
$S'\rightarrow S$ is a B-move. The assertion has been proven.

{\sl Example:} Here is a path from an A-string
to a B-string:
\def\X(#1,#2){L^{#2}_{-#1}}
\[
\begin{array}{l}
\X(4,+)\X(4,+)\X(3,+)\X(2,-)\X(1,+)\X(1,+)
{}~\longrightarrow   \\
\X(4,+)\X(3,+)\X(2,+)\X(2,-)\X(2,-)\X(1,+)\X(1,+)
 ~\longrightarrow  \\
\X(3,+)\X(2,+)\X(2,+)\X(2,-)\X(2,-)\X(2,-)\X(1,+)\X(1,+)
 ~\longrightarrow  \\
\X(3,+)\X(2,+)\X(2,-)\X(2,-)\X(2,-)\X(1,+)\X(1,+)\X(1,+)\X(1,-)
 ~\longrightarrow  \\
\X(3,+)\X(2,-)\X(2,-)\X(2,-)\X(1,+)\X(1,+)\X(1,+)\X(1,+)\X(1,-)\X(1,-)~.
\end{array}
\]

\setcounter{section}{3}
\setcounter{equation}{0}

\section{Vertex operators}

We now define type-I and type-II vertex operators as intertwiners of
certain $\aqp$-modules, under the assumption of the existence and
uniqueness of such vertex operators. To minimize the number of
necessary assumptions, we start by defining type-I vertex operator
$\Phi(\z)$, and then construct type-II vertex operator $\Psi^*(\z)$ in
terms of $L(\z)$ and $\Phi(\z)$. These operators are deformations of
the corresponding operators in the trigonometric case, which are
reviewed in Appendix A.

\subsection{Type-I vertex operators}

Let $$s=q^{2(k+2)},\quad y_{l}^+=q^{2(l+1)},\quad y_l^-=s
(y_l^+)^{-1},$$ and let $L^{(l)}(\zeta)$ denote $L(\z)$ acting on
$\M_{k,l}$.  We expect the following to hold for generic values of $k$
and $l$:

\noindent(i)
There exists a unique, up to normalization, intertwiner of
$\aqp$-modules, the type-I vertex operator $\Phi^{(l\pm1,l)}(\z)$, $$
\Phi^{(l\pm 1,l)}(\z)
: \M_{k,l} \longrightarrow \M_{k,l\pm 1}\otimes V_\z,
\quad \Phi^{(l\pm 1,l)}(\z)
=\sum_{\vep}\Phi_\vep^{(l\pm 1,l)}(\z)\otimes v_\vep~,
$$
satisfying the intertwining relation
\begin{equation}
\Phi^{(l\pm 1,l)}_{\vep_2}(\z_2) L^{(l)}_{\vep_1 \vep_1''}(\z_1)=
\sum_{\vep_1'\vep_2'}R^+_{\vep_1 \vep_2;\vep_1'\vep_2'}(\z_1/\z_2)
L^{(l\pm 1)}_{\vep_1'\vep_1''}(\z_1)\Phi^{(l\pm 1,l)}_{\vep_2'}(\z_2)~.
\label{LPhi}
\eeq
The Laurent components of $\Phi^{(l',l)}_\e(\z)$ are defined as
\beq\label{components}
\Phi^{(l\pm 1,l)}_\vep(\z)=\sum_{n\equiv\frac{1\pm\e}{2}(\mod 2)}
\varphi^{(l\pm 1,l)}_{n}\z^{-n},
\eeq
where
$$\varphi^{(l\pm 1,l)}_{n}:\bigl(\M_{k,l}\bigr)_d\rightarrow
\bigl(\M_{k,l\pm1}\bigr)_{d-n},$$
and we choose to normalize the action on $\MKL$ by
\beq
\varphi^{(l\pm 1,l)}_{0}|k,l\rangle=|k,l\pm1\rangle~.\label{phinormalization}
\eeq
By convention,  $\Phi_\vep^{(l',l)}(\z)=0$ if $l'\neq l\pm1$.

\noindent(ii)
The type-I vertex operators satisfy the commutation relations
\begin{eqnarray}
&&R_{12}(\zeta_1/\zeta_2)
{\Phim{1}}^{(l'',l')}(\zeta_1)
{\Phim{2}}^{(l',l)}(\zeta_2)
\nonumber \\
&& \qquad =\sum_{\sigma=\pm 1}
w\left(\matrix{\phantom{l''}l'\phantom{l}\cr
                   l''\phantom{l+\sigma}l \cr
               \phantom{l''}l+\sigma\phantom{l}\cr}
\Biggl|\zeta_1/\zeta_2\right)
{\Phim{2}}^{(l'',l+\sigma)}(\zeta_2)
{\Phim{1}}^{(l+\sigma,l)}(\zeta_1),
\label{PhiPhi}
\end{eqnarray}
where the connection coefficients $w$ are defined in (\ref{w}), and
the matrix $R$ is defined in equation (\ref{R}).  This should be
understood as a relation among matrix elements (cf. the remark at the
end of Appendix A).

\noindent(iii)
The vertex operators satisfy the inversion relation
\begin{eqnarray}
g_{l',l}\sum_\vep
\Phi^{(l,l')}_{-\vep}(-q^{-1}\zeta)
\Phi^{(l',l)}_{\vep}(\zeta)
&=&{\rm id},
\label{inv1}
\end{eqnarray}
or, equivalently,
\begin{eqnarray}
\sum_{l'} g_{l,l'}
\Phi^{(l,l')}_{\vep_1}(\zeta)
\Phi^{(l',l)}_{-\vep_2}(-q^{-1}\zeta)
&=&
\delta_{\vep_1,\vep_2}{\rm id},
\label{inv2}
\end{eqnarray}
where
\begin{eqnarray}\label{g-def}
&&g_{l\pm1,l}
=
\frac{(y^\pm_l;s)_\infty}{(q^2y^\pm_l;s)_\infty}
\frac{\widetilde{\xi}(p;p,q)}{\widetilde{\xi}(s;s,q)}, \\
&&\widetilde{\xi}(z;p,q)
=\frac{(q^2z;p,q^4)^2_\infty}{(z;p,q^4)_\infty(q^4z;p,q^4)_\infty}
=\xi(z;p,q)\frac{(z;q^4)_\infty}{(q^2z;q^4)_\infty}.
\end{eqnarray}
The scalar $g_{l',l}$ is determined by taking the expectation values
of (\ref{inv1}) (see Appendix B).

\medskip

Note that the connection coefficients $w$ in (\ref{PhiPhi}) are the
same as in the trigonometric case (see Appendix A), though {\it a
priori} they might depend on $p$. A word is in order concerning this
point in the elliptic case.  For the consistency of (\ref{PhiPhi}),
the coefficients $w$ must satisfy the Yang-Baxter equation. It seems
unlikely that $w$, which is already an elliptic solution, extends to a
more general class of solutions.  Thus we expect $w$ to remain
unchanged under the elliptic deformation.

To show the equivalence of (\ref{inv1}) and (\ref{inv2}),
we use (\ref{SPVL}), as well as the relation
$$
w\left(\matrix{\phantom{l}l'\phantom{l}\cr
                  l\phantom{l''}l \cr
               \phantom{l}l''\phantom{l}\cr}
\Biggl|-q^{-1} \right)
={g_{l,l''}\over g_{l',l}}~.
$$
{}From (\ref{inv1}), we have
\begin{eqnarray*}
\delta_{\vep_1,\vep_2}{\rm id}
&=&
g_{l',l}\delta_{\vep_1,\vep_2}
\sum_{\vep}\Phi^{(l,l')}_{-\vep}(-q^{-1}\zeta)
           \Phi^{(l',l)}_{\vep}(\zeta)
\\
&=&
g_{l',l}
\sum_{\vep}R_{\vep_1,-\vep_2;-\vep,\vep}(-q^{-1})
\Phi^{(l,l')}_{-\vep}(q^{-1}\zeta)
\Phi^{(l',l)}_{\vep}(\zeta)
\\
&=&
\sum_{l''}g_{l',l}
w\left(\matrix{\phantom{l}l'\phantom{l}\cr
                  l\phantom{l''}l \cr
               \phantom{l}l''\phantom{l}\cr}
\Biggl|-q^{-1} \right)
\Phi^{(l,l'')}_{\vep_1}(\zeta)
\Phi^{(l'',l)}_{-\vep_2}(-q^{-1}\zeta)
\\
&=&
\sum_{l'}g_{l,l'}
\Phi^{(l,l')}_{\vep_1}(\zeta)
\Phi^{(l',l)}_{-\vep_2}(-q^{-1}\zeta).
\end{eqnarray*}

\subsection{Type-II vertex operators}

We now define type-II vertex operators $\Psi^*(\z)$ in terms of the
$L$-operator and type-I vertex operators.  We show that $\Psi^*(\z)$
thus defined satisfy the intertwining
relation (\ref{LPsi}) and the commutation relations (\ref{PhiPsi}),
(\ref{PsiPsi}).

The type-II vertex operator is an intertwiner of $\aqp$-modules
\[
\Psi^{*(l\pm 1,l)}(\z)
:V_\z\otimes \M_{k,l} \longrightarrow \M_{k,l\pm 1},~~
 \Psi^{*(l\pm 1,l)}_\vep(\z)
= \Psi^{*(l\pm 1,l)}(\z)\left(v_\vep\otimes\cdot\right).
\]
It can be defined in the following way:
\beq
{\Psi}^{*(l,l')}_{\vep'}(\zeta)=c_{l,l'}\sum_{\vep}
L^{(l)}_{\vep \vep'}(q^{k/2}\zeta)
 \Phi^{(l,l')}_{-\vep}(-q^{k/2+1}\zeta)~,
\label{Psi}
\eeq
where
\begin{equation}\label{c-const}
c_{l,l\pm 1}=a^{-1}\widetilde{\xi}(p;p,q)\overline{c}_{l,l\pm 1},
\quad
\overline{c}_{l,l\pm1}=\cases{q^{-l/2}&for $+$;\cr q^{-(k-l)/2}&for $-$.\cr}
\en
Then, as we show below, it follows that $\Psi^*$ satisfies the
intertwining relation
\begin{eqnarray}
&&L^{(l)}_{\vep_2\vep_2''}(\z_2)\Psi^{*(l,l\pm 1)}_{\vep_1''}(\z_1)\nonumber\\
&&\qquad=\sum_{\vep_1',\vep_2'}
\Psi^{*(l,l\pm 1)}_{\vep_1'}(\z_1)L^{(l\pm 1)}_{\vep_2\vep_2'}(\z_2)
R^{*+}_{\vep_2'\vep_1';\vep_2''\vep_1''}(q^{-k/2}\z_2/\z_1)~.
\label{LPsi}
\end{eqnarray}
The components of $\Psi^{*(l',l)}_\e(\z)$ are defined as
$$
\Psi^{*(l\pm 1,l)}_\vep(\z)=\sum_{n\equiv\frac{1\mp\e}{2}(\mod 2)}
 \psi^{*(l\pm 1,l)}_{n}\z^{-n},$$
where
$$
\psi^{*(l\pm 1,l)}_{n}:(\M_{k,l})_d\longrightarrow (\M_{k,l\pm 1})_{d-n},
$$
and we have chosen the scalar $c_{l,l'}$ in (\ref{Psi}) to ensure the
normalization
\be
\psi^{*(l\pm 1,l)}_{0}\ket{k,l}=\ket{k,l\pm 1}\label{PSNR}
\eeq
(see Appendix B for a detailed derivation).

For the case of level-one, the relation (\ref{Psi}) has been presented in
\cite{FIJKMY}.

In addition, it is possible to show that the type-II vertex operators satisfy
the following commutation relations:
\begin{eqnarray}
&&
{\Phim{1}}^{(l'',l')}(\zeta_1)
{\Psim{2}}^{(l',l)}(\zeta_2)
\nonumber \\
&&\quad =
\sum_{\sigma=\pm 1}
\widetilde{w}\left(\matrix{\phantom{l''}l'\phantom{l}\cr
                   l''\phantom{l+\sigma}l \cr
               \phantom{l''}l+\sigma\phantom{l}\cr}
\Biggl|\zeta_1/\zeta_2\right)
{\Psim{2}}^{(l'',l+\sigma)}(\zeta_2)
{\Phim{1}}^{(l+\sigma,l)}(\zeta_1),
\label{PhiPsi} \\
&&
-{\Psim{1}}^{(l'',l')}(\zeta_1)
 {\Psim{2}}^{(l',l)}(\zeta_2)
\nonumber \\
&&\quad=
\sum_{\sigma=\pm 1}
w\left(\matrix{\phantom{l''}l'\phantom{l}\cr
            l''\phantom{l+\sigma}l \cr
            \phantom{l''}l+\sigma\phantom{l}\cr}
\Biggl|\zeta_1/\zeta_2\right)
\Psim{2}^{(l'',l+\sigma)}(\zeta_2)
\Psim{1}^{(l+\sigma,l)}(\zeta_1)R^*_{12}(\zeta_1/\zeta_2),\nonumber\\
\label{PsiPsi}
\end{eqnarray}
where the connection coefficients $\widetilde{w}$ are defined in (\ref{wtil}).

To show that the intertwining relation (\ref{LPsi}) follows from the
definition (\ref{Psi}), we use (\ref{LPhi}), (\ref{PhiPhi}),
(\ref{RLL}), (\ref{crossing}) and (\ref{unitarity}) to obtain
\begin{eqnarray*}
&&\sum_{\vep_1',\vep_2'}\Psi^{*(l,l')}_{\vep_1'}(\zeta_1)
L^{(l')}_{\vep_2\vep_2'}(\zeta_2)
R^{*+}_{\vep_2'\vep_1';\vep_2''\vep_1''}(q^{-k/2}\zeta_2/\zeta_1) \\
&&=
\sum_{\vep_1,\vep_1',\vep_2'}c_{l,l'}
L^{(l)}_{\vep_1\vep_1'}(q^{k/2}\zeta_1)
\Phi^{(l,l')}_{-\vep_1}(-q^{k/2+1}\zeta_1)
L^{(l')}_{\vep_2\vep_2'}(\zeta_2)\\&& \qquad\qquad\times
R^{*+}_{\vep_2'\vep_1';\vep_2''\vep_1''}(q^{-k/2}\zeta_2/\zeta_1) \\
&&=
\sum_{\vep_1,\vep_1',\vep_2',\sigma_1,\sigma_2}c_{l,l'}
L^{(l)}_{\vep_1\vep_1'}(q^{k/2}\zeta_1)
R^{+}_{\vep_2-\vep_1;\sigma_2-\sigma_1}(-q^{-k/2-1}\zeta_2/\zeta_1)
L^{(l)}_{\sigma_2\vep_2'}(\zeta_2) \\
&& \qquad\qquad\times
\Phi^{(l,l')}_{-\sigma_1}(-q^{k/2+1}\zeta_1)
R^{*+}_{\vep_2'\vep_1';\vep_2''\vep_1''}(q^{-k/2}\zeta_2/\zeta_1) \\
&&=
\sum_{\vep_1,\vep_1',\vep_2',\sigma_1,\sigma_2}c_{l,l'}
R^{+}_{\vep_2-\vep_1;\sigma_2-\sigma_1}(-q^{-k/2-1}\zeta_2/\zeta_1)
R^{+}_{\sigma_2\vep_1;\vep_2'\vep_1'}(q^{-k/2}\zeta_2/\zeta_1) \\
&& \qquad\qquad\times
L^{(l)}_{\vep_2'\vep_2''}(\zeta_2)
L^{(l)}_{\vep_1'\vep_1''}(q^{k/2}\zeta_1)
\Phi^{(l,l')}_{-\sigma_1}(-q^{k/2+1}\zeta_1)\\
&&=\sum_{\vep_1,\vep_1',\vep_2',\sigma_1,\sigma_2}
R^{-}_{\sigma_1\vep_2;\vep_1\sigma_2}(q^{k/2}\zeta_1/\zeta_2)
R^{+}_{\sigma_2\vep_1;\vep_2'\vep_1'}(q^{-k/2}\zeta_2/\zeta_1)
L^{(l)}_{\vep_2'\vep_2''}(\zeta_2) \\
&&\qquad\qquad \times
c_{l,l'}
L^{(l)}_{\vep_1'\vep_1''}(q^{k/2}\zeta_1)
\Phi^{(l,l')}_{-\sigma_1}(-q^{k/2+1}\zeta_1) \\
&&=
L^{(l)}_{\vep_2'\vep_2''}(\zeta_2)
\Psi^{*(l,l')}_{\vep_1''}(\zeta_1). \qquad
\end{eqnarray*}

Similarly, to prove (\ref{PhiPsi}), we use (\ref{Psi}), (\ref{LPhi}),
(\ref{SPVL}) and (\ref{PhiPhi}).  For (\ref{PsiPsi}), we use
(\ref{Psi}), (\ref{PhiPsi}), (\ref{LPsi}) and the fact that
\[
\frac{c_{l'',l'}c_{l',l}}
     {c_{l'',l+\sigma}c_{l+\sigma,l}}
=1 \quad \mbox{if} \quad
{w}\left(\matrix{\phantom{l''}l'\phantom{l}\cr
                  l''\phantom{l+\sigma}l \cr
               \phantom{l''}l+\sigma\phantom{l}\cr}
\Biggl|\z \right)
\neq 0.
\]

Finally we
note that the inversion relation (\ref{inv2}) enables us to express
$L^{(l)}(\z)$ in terms of type-I and type-II vertex operators
(cf. (\ref{trigLPsiPhi})):
\begin{eqnarray*}
L^{(l)}_{\vep,\vep'}(\zeta)
&=&
\sum_{\sigma=\pm 1}\frac{g_{l,l+\sigma}}{c_{l,l+\sigma}}
{\Psi^*}^{(l,l+\sigma)}_{\vep'}(q^{-k/2}\zeta)
 \Phi^{(l+\sigma,l)}_{\vep}(\zeta).
\end{eqnarray*}
This is a generalization of the relation pointed out by Miki for the
trigonometric case at level-one~\cite{Miki}.

\setcounter{section}{4}
\setcounter{equation}{0}

\section{Ordering-rules for $\Phi$}

The defining relation (\ref{LPhi}) of the type-I vertex operators
$\Phi(\z)$ can be written as an ordering rule between the Laurent
components of $\Phi$ and $L$. This ordering rule determines the action
of $\Phi_\e^{(l\pm1,l)}(\z)$: $\M_{k,l}\rightarrow\M_{k,l\pm1}$, as we
explain below.

Alternatively, the components $\varphi_n^{(l\pm1,l)}$ (cf.
(\ref{components})) can be used to create vectors in $\M_{k,l}$ in a
similar manner as the operators $L^{(l)\e}_{n}$, with the defining
relation of the algebra (\ref{RLL}) replaced by the commutation
relation between $\Phi$-operators (\ref{PhiPhi}). However, whereas the
$L$ operator acts on $\M_{k,l}$ to create a basis from a single
highest weight vector $|k,l\rangle$, strings of operators
$\varphi_n^{(l,l\pm1)}$ create vectors in $\M_{k,l}$ from vectors in
$\M_{k,l+j},~j\in\Z$.  Such a construction of the vectors in highest
weight module has not been presented before, even in the trigonometric
case.

\subsection{Ordering-rules for $L(\z)$ and $\Phi(\z)$}

To write down the ordering-rules for $\varphi_m$ and $L^\e_{n}$, let
\begin{equation}\label{ab-func}
\alpha^{(1)\sigma}(\z)=h^{(1)\sigma}(\z),\qquad
\alpha^{(2)\sigma}(\z)=\rho(\z^2)^{-1}h^{(2)\sigma}(\z)
\end{equation}
with $\rho(\z^2)$ defined in (\ref{rho}) and $h^{(i)\sigma}(\z)$ in
(\ref{h-def}).  The functions $\alpha^{(i)\sigma}(\z)$ are formal
power series in the variable $\z$, and we rewrite (\ref{LPhi}) as an
equality of formal power series:
\begin{equation}\label{lphi-split}
\alpha^{(2)\sigma}(\z_1/\z_2)\sum_{\vep_1=\pm}
\Phi_{\sigma \vep_1}^{(l,l')}(\z_2)
L^{(l')}_{\vep_1\vep_1''}(\z_1) =
\alpha^{(1)\sigma}(\z_2/\z_1)\sum_{\vep_1'=\pm}L^{(l)}_{\vep_1'\vep_1''}(\z_1)
\Phi_{\sigma\vep_1'}^{(l,l')}(\z_2),
\end{equation}
where $\sigma=\pm$. In a similar manner as in Section 3.2, this
equation implies the following ordering-rules:
\begin{eqnarray}\label{lphiorder}
L_{n}^{(l)\e}\varphi^{(l,l')}_{m}
&=&
\sum_{j\geq 0}
\varphi_{m+n-j}^{(l,l')}L_{j}^{(l')\e}
\sum_{k=0}^j
\widetilde{\alpha}^{(1)\sigma}_{n-j+k}\alpha^{(2)\sigma}_k\nonumber
\\
&&\quad +
\sum_{j>0}
L_{-j}^{(l)\e}\varphi_{m+n+j}^{(l,l')}
\sum_{k=0}^{j-1}
\widetilde{\alpha}^{(1)\sigma}_{n+j-k}\alpha^{(1)\sigma}_k ~,\nonumber
\\
\varphi_{m}^{(l,l')}L_{-n}^{(l')\e}
&=&
\sum_{j>0}
L_{-j}^{(l)\e}\varphi_{m-n+j}^{(l,l')}
\sum_{k=0}^{j-1}
\widetilde{\alpha}^{(2)\sigma}_{n-j+k}\alpha^{(1)\sigma}_k\nonumber
\\
&&\quad +
\sum_{j\geq0}
\varphi_{m-n-j}^{(l,l')}L_{j}^{(l')\e}
\sum_{k=0}^j
\widetilde{\alpha}^{(2)\sigma}_{n+j-k}\alpha^{(2)\sigma}_k,
\end{eqnarray}
where $\sigma=(-1)^{m-n}(l'-l)\vep$, and the Taylor coefficients are
defined as
$$\alpha^{(i)\sigma}(\z)=\sum_{j\geq0}\alpha^{(i)\sigma}_j\z^j=
{1\over\sum_{j\geq0}
\widetilde{\alpha}^{(i)\sigma}_j\z^j}~.$$

Equations (\ref{lphiorder}) determine the action of
$\varphi_n^{(l\pm1,l)}$ on the basis (\ref{LBASIS}) generated by $L$,
in the following sense.  The action of $\varphi_n^{(l\pm1,l)}$ is
determined if we give all the matrix elements of the form $\langle
w|\varphi_n|w'\rangle$ where $\langle w|\in\M^*_{k,l\pm1}$ and
$|w'\rangle\in\MKL$.  This is computable by using the ordering rules
(\ref{lphiorder}): We can use (\ref{lphiorder}) to reduce the
calculation of $\langle w|\varphi_n|w'\rangle$ to that for $\langle
w|$ and $|w'\rangle$ with lower degrees.  Note that the ordering rules
(\ref{lphiorder}) do not lower the degree if $n\le0$, but are
trivially true in that case.

\subsection{Normal-ordering of $\Phi$ operators}

A conjectural basis for $\MKL$ can be generated by the action of the
operators $\varphi_m^{(l\pm1,l)}$ as follows.  We first compute the
normal-ordering of products of the form
$\varphi_m^{(l,l')}\varphi_n^{(l',l'')}$, by again rewriting the
commutation relation (\ref{PhiPhi}) as a relation between formal power
series:
\begin{eqnarray}
&&\alpha_1(\zeta_2/\zeta_1)
\sum_{m\equiv n~(2)}
\varphi^{(l\pm1,l)}_m\varphi^{(l,l\mp1)}_n \zeta_1^{-m}\zeta_2^{-n}=
(\zeta_1\leftrightarrow \zeta_2),\label{phiphi1}\\ &&
\frac{\beta_1(\zeta_2/\zeta_1)}{1+q^{-1}\zeta_2/\zeta_1}
\sum_{m\not \equiv n~(2)}
\varphi^{(l\pm1,l)}_m\varphi^{(l,l\mp1)}_n \zeta_1^{-m}\zeta_2^{-n}
\nonumber\\ & &\mbox{\hspace{3cm}}=
q\frac{\beta_1(\zeta_1/\zeta_2)}{1+q\zeta_1/\zeta_2}
\sum_{m\not \equiv n~(2)}
\varphi^{(l\pm1,l)}_n\varphi^{(l,l\mp1)}_m \zeta_2^{-n}\zeta_1^{-m},
\label{phiphi2}\\ &&
\sum_{m\not \equiv n~(2)}
\varphi^{(l)}(m,n)\zeta_1^{-m-1}\zeta_2^{-n}
\alpha_1(\zeta_2/\zeta_1)F^{(l)}(\zeta_2/\zeta_1)^{-1}=
(\zeta_1\leftrightarrow \zeta_2),\\&&
\sum_{m\equiv n~(2)}
\varphi^{(l)}(m,n)\zeta_1^{-m}\zeta_2^{-n}
\beta_1(\zeta_2/\zeta_1)F^{(l)}(\zeta_2/\zeta_1)^{-1}
=(\zeta_1\leftrightarrow \zeta_2),
\end{eqnarray}
where  $\varphi^{(l)}(m,n)=(\varphi^{(l,l+1)}_m\varphi^{(l+1,l)}_n,
\varphi^{(l,l-1)}_m\varphi^{(l-1,l)}_n),$
\[
\alpha_1(\zeta)=\alpha(\zeta)\xi(\zeta^2;s,q)
\qquad\beta_1(\zeta)=\beta(\zeta)\xi(\zeta^2;s,q),
\]
\beqa
&&a(\zeta)+d(\zeta)=\zeta^{-1}\frac{\alpha(\zeta^{-1})}{\alpha(\zeta)},
\quad
\alpha(\zeta)=
\frac{(p^{1/2}q\zeta;p)_\infty}{(p^{1/2}q^{-1}\zeta;p)_\infty}
\xi(\zeta^2;p,q)^{-1},
\nonumber\\
&&b(\zeta)+c(\zeta)=\frac{\beta(\zeta^{-1})}{\beta(\zeta)},
\qquad
\beta(\zeta)=\frac{(-q\zeta;p)_\infty}{(-pq^{-1}\zeta;p)_\infty}
\xi(\zeta^2;p,q)^{-1}.
\label{alphabeta}
\eeqa
The function $\xi(z;p,q)$ is defined in (\ref{xi}).  The
matrix $F^{(l)}(\z)$ is related to the Riemann-Hilbert splitting of
the matrix $W$, and corresponds to the connection formula for the
two-point function of the $q$KZ equation~\cite{FreResh}:
\[
W^{(l)}(\z)=
\left(\matrix{w^{(l)}_{++}(\z) & w^{(l)}_{+-}(\z) \cr
              w^{(l)}_{-+}(\z) & w^{(l)}_{--}(\z) \cr}\right)
=F^{(l)}(\z)^{-1}\,F^{(l)}(\z^{-1}),
\]
where $w$ are defined in (\ref{wpm}) and
\begin{eqnarray}
&&F^{(l)}(\zeta)=
\left(\matrix{  f_{++}^{(l)}(\zeta)  & f_{+-}^{(l)}(\zeta) \cr
                f_{-+}^{(l)}(\zeta) & f_{--}^{(l)}(\zeta) \cr}\right),\nonumber
\\
&&F^{(l)}(\zeta)^{-1}=
\frac{(s q^{-2}\zeta^2;s)_\infty}{(q^2\zeta^2;s)_\infty}
\left(\matrix{  f_{--}^{(l)}(\zeta)  & -f_{+-}^{(l)}(\zeta) \cr
             -f_{-+}^{(l)}(\zeta)  & f_{++}^{(l)}(\zeta) \cr}\right),\nonumber
\\
&&
f_{\pm\pm}^{(l)}(\zeta)=
{}_2\phi_1\left(\matrix
{q^2 \phantom{y^\pm_l} s^{-1}q^2y^\pm_l \cr
\phantom{q^2}y^\pm_l\phantom{s^{-1}q^2y^\pm_l}\cr}
;s,s q^{-2}\zeta^2\right),\nonumber
\\
&&f_{\mp\pm}^{(l)}(\zeta)=
\frac{y^\pm_l-q^2}{1-y^\pm_l}(-q)^{-1}\zeta\times
{}_2\phi_1\left(\matrix
{q^2 \phantom{s y^\pm_l} q^2y^\pm_l \cr
\phantom{q^2}s y^\pm_l\phantom{q^2y^\pm_l}\cr}
;s,s q^{-2}\zeta^2\right).\label{f-def}
\end{eqnarray}
The basic hypergeometric function ${}_2\phi_1$ is defined in
(\ref{hyper}).

We remark that the choice of Riemann-Hilbert splitting in equation
(\ref{phiphi2}) is not unique. We choose this factorization since it
is the correct relation at $p=0$ and it agrees with checks performed
for low degree matrix elements. The ordering rules below are invariant
under the interchange of $\z_1$ and $\z_2$ in equation
(\ref{phiphi2}). For further discussion of this point the reader is
referred to Appendix A.

With this choice, the normal-ordering rules are now computed in a
straightforward manner to be
\begin{eqnarray}
\varphi^{(l\pm 1,l)}_m\varphi^{(l,l\mp 1)}_n
&&=
\sum_{i\ge 0} e_{m-n,i}
\varphi^{(l\pm 1,l)}_{\left[\frac{m+n}{2}\right]_- -i}
\varphi^{(l, l\mp 1)}_{\left[\frac{m+n}{2}\right]_+ +i}\nonumber
\\
\varphi^{(l)}(m,n)
&&=
\sum_{i\ge 0}
\varphi^{(l)}({\scriptstyle{\left[\frac{m+n}{2}\right]_- -i,~
\left[\frac{m+n}{2}\right]_+ +i}})
E^{(l)}_{m-n,i},
\label{Phinormal}
\end{eqnarray}
where
\begin{eqnarray}
e_{m,i}
&=&
\sum_{|k|\le i} \alpha_{1,i-|k|}\widetilde{\alpha}_{1,\frac{m}{2}+k}
\quad (m \hbox{ even}),\nonumber
\\
&=&
\sum_{0\le k\le i}
\left(\beta_{2,i-k}\widetilde{\beta}_{1,\frac{m-1}{2}-k}
+\beta_{1,i-k}\widetilde{\beta}_{1,\frac{m+1}{2}+k}\right)
\quad (m \hbox{ odd}),\nonumber
\\
E^{(l)}_{m,i}
&=&
\sum_{|k|\le i} B^{(l)}_{i-|k|}\widetilde{B}^{(l)}_{\frac{m}{2}+k}
\quad (m \hbox{ even}),\nonumber
\\
&=&
\sum_{|k|\le i} A^{(l)}_{i-|k|}\widetilde{A}^{(l)}_{\frac{m+1}{2}+k}
\quad (m \hbox{ odd}),
\label{eE}
\end{eqnarray}
with the coefficients defined by
\begin{eqnarray*}
&&\alpha_1(\zeta)=\sum_{j\ge 0}\alpha_{1,j}\zeta^j,\quad
\frac{\beta_1(\zeta)}{1+q^{-1}\zeta}
=\sum_{j\ge 0}\beta_{1,j}\zeta^j,
\quad
\frac{q\beta_1(\zeta)}{1+q \zeta}
=\sum_{j\ge 0}\beta_{2,j}\zeta^j,
\\
&&\alpha_1(\zeta)F^{(l)}(\zeta)^{-1}
=
\sum_{j\ge 0} A^{(l)}_j\zeta^j,
\quad
\beta_1(\zeta)F^{(l)}(\zeta)^{-1}
=
\sum_{j\ge 0} B^{(l)}_j\zeta^j.
\end{eqnarray*}
and
$\sum\widetilde{\alpha}_{1,j}\zeta^j=\alpha_1(\zeta)^{-1}$, etc..

Using the inversion relation (\ref{inv1}), we can also rewrite
products of the form $\varphi_n^{(l,l\pm1)}\varphi_n^{(l\pm1,l)}$ as a
linear combination of normal-ordered products:
\beq
\varphi^{(l)}(n,n) = \delta_{n,0}(1,1) -
\Bigl(\sum_{m>0}\varphi^{(l)}({\scriptstyle {n-m,n+m}}) D^{(l)}_m\Bigr)
\beta_1(-q)F^{(l)}(-q)^{-1},
\label{inversion}
\eeq
where
\[
D_m^{(l)} =(-q)^{-m} {\rm id} + \sum_{k>0}(-q)^k E_{2k,m}^{(l)}
\]
and $g_{l\pm1,l}$ is defined in (\ref{g-def}).

\subsection{Vectors generated by $\Phi$-operators}

We now specify a set of vectors in $\M_{k,l}$ as follows.  Let
$\{n_j\}_{j=1}^{m}$ be a sequence of negative integers, and let
$\{l_j\}_{j=0}^m$, $l_0=l$, be a corresponding sequence with
$l_i-l_{i-1}=\pm1$.  Then we conjecture that $\M_{k,l}$ is spanned by
the set of basis vectors of the form
\begin{eqnarray}
&\varphi_{n_1}^{(l,l_1)}\varphi_{n_2}^{(l_1,l_2)}\cdots
\varphi_{n_m}^{(l_{m-1},l_m)}|k,l_m\rangle,&\nonumber\\
&n_1\leq n_2\leq \cdots\leq n_m,&\nonumber\\
&\hbox{if}~~n_i=n_j~~\hbox{then}~~l_i-l_{i-1}=l_j-l_{j-1}~.&\label{basis}
\end{eqnarray}

The normal-ordering rules (\ref{Phinormal}) and
(\ref{inversion}) allow us to express all other $\varphi$-strings as
linear combinations of vectors of the form (\ref{basis}).

The vectors (\ref{basis}) have a natural grading with degree
$d=-\sum n_j$. The number of degree-$d$ basis vectors of the form
(\ref{basis}) is equal to the number of degree-$d$ $L$-basis vectors,
when $(k,l)$ are generic.
However, degree-$d$ $\varphi$-basis
vectors are created from a maximum of $2d+1$ highest weight vectors,
whereas the $L$-basis is created from a single highest weight vector.

Similarly we conjecture that  $\M^*_{k,l}$ is spanned by
left vectors of the form
\begin{equation}
\langle k,l_m|\varphi_{-n_m}^{(l_m,l_{m-1})}\cdots
\varphi_{-n_2}^{(l_{2},l_1)}
\varphi_{-n_1}^{(l_{1},l_0)}\label{dualbasis}
\end{equation}
with degree $d=-\sum n_j$. In Appendix C, we show that the
Shapovalov-type determinant $\Delta_d$ of the matrix of couplings of
degree-$d$ vectors of the form (\ref{basis}) in $\MKL$ with those of
the form (\ref{dualbasis}) in $\DMKL$, has a simple factorized form
for low degree, similar to the determinant computed for the components
of the $L$-operator.

\setcounter{equation}{0}
\setcounter{section}{5}

\section{Level-one case}
The vertex operators for the level-one irreducible highest weight
modules for $\aqp$ were discussed in \cite{FIJKMY}.  In this section
we start from the commutation relation and the inversion relation for
the type-I vertex operator $\Phi_\varepsilon^{(1-i,i)}(\z):{\cal
H}^{(i)}\rightarrow{\cal H}^{(1-i)},~ (i=0,1$) conjectured in \cite{FIJKMY}.
\be
R(\z_1/\z_2)\Phim{1}(\z_1)\Phim{2}(\z_2)
=
\Phim{2}(\z_2)\Phim{1}(\z_1)~,\lb{EQUAA}
\en
\be
g\sum_\varepsilon\Phi_{-\varepsilon}^{(i,1-i)}(-q^{-1}\z)
\Phi_\varepsilon^{(1-i,i)}(\z)={\rm id}.\label{EQUAB}
\en
Note that the module ${\cal H}^{(i)}$ is graded:
${\cal H}^{(i)}=\oplus_{d=0}^\infty {\cal H}^{(i)}_d$,
and is generated from the highest weight vector $|i\rangle$.
We conjecture that the module ${\cal H}^{(0)}\oplus{\cal H}^{(1)}$
is generated from $|0\rangle$ and $|1\rangle$ by the actions
of $\varphi^{(1-i,i)}_n$ $(n<0)$.

Let us write the ordering rules among the $\varphi^{(1-i,i)}_n$.
We choose the Riemann-Hilbert splitting as follows (see Section 2.1
for the definition of $R$):
\begin{eqnarray*}
&&
\z_1^{-1}\alpha(\zeta_2/\zeta_1)
\sum_{m\not \equiv n~(2)}
\varphi^{(i,1-i)}_m\varphi^{(1-i,i)}_n \zeta_1^{-m}\zeta_2^{-n}
=
(\zeta_1\leftrightarrow \zeta_2),
\\
&&
\beta(\zeta_2/\zeta_1)
\sum_{m \equiv n~(2)}
\varphi^{(i,1-i)}_m\varphi^{(1-i,i)}_n \zeta_1^{-m}\zeta_2^{-n}
=
(\zeta_1\leftrightarrow \zeta_2),
\end{eqnarray*}
where $\alpha(\z)$ and $\beta(\z)$ are given by (\ref{alphabeta}).
Set
\begin{eqnarray*}
&&\alpha(\z)=\sum_{j\ge 0}\alpha_j \z^j
={1\over\sum_{j\ge 0}\widetilde{\alpha}_j \z^j},\\
&&\beta(\z)=\sum_{j\ge 0}\beta_j \z^j=
{1\over\sum_{j\ge 0}\widetilde{\beta}_j \z^j}.
\end{eqnarray*}
{}From \refeq{EQUAA} we obtain the normal-ordering rules
\be
\varphi_m^{(i,1-i)}\varphi_n^{(1-i,i)}=
\displaystyle \sum_{j\ge 0}\displaystyle \varphi^{(i,1-i)}
_{\frac{m+n-\delta}{2}-j}
\varphi^{(1-i,i)}_{\frac{m+n+\delta}{2}+j} d_{m-n,j}~ ,~~m>n,\\
\en
where
\[
\delta=\cases{0&if $m-n$ is even;\cr1&if $m-n$ is odd,\cr}
\]
and
\[
d_{m,j}=\cases{
\displaystyle \sum_k\displaystyle \widetilde{\beta}_{{m\over 2}-k}\beta_{j-|k|}
&if $m$ is even;\cr
\displaystyle \sum_k\displaystyle \widetilde{\alpha}_{{m+1\over
2}-k}\alpha_{j-|k|}
&if $m$ is odd.\cr}
\]
In addition, from (\ref{EQUAB}) we have
\[
\varphi_m^{(i,1-i)}\varphi_m^{(1-i,i)}
=\delta_{m0}-\sum_{j\ge1}\varphi_{m-j}^{(i,1-i)}\varphi^{(1-i,i)}_{m+j}
\sum_{k}\beta_{j-|k|}(-q)^k.
\]

By using these ordering rules our conjecture is refined to
the statement that ${\cal H}^{(i)}_d$ is spanned by the vectors
\be
\varphi_{n_1}^{(i,i+1)}\cdots\varphi_{n_m}^{(i+m-1,i+m)}|i+m\rangle
\label{basis1}
\en
such that
\[
n_1<\cdots<n_m<0\hbox{~~ and~~ }-\sum_{j=1}^mn_j=d.
\]
The number $N_d$ of such vectors are given by the generating function
\[
\sum_{d=0}^\infty N_d~t^d=\prod_{n=1}^\infty(1+t^n).
\]
This is equal to the character of the level-one ${\widehat{sl}}_2$-modules.
Therefore, we further conjecture that these vectors constitute a basis.

As in the generic level case of in Sections 3 and 5, the
Shapovalov-type determinant factorizes for small $d$, as we discuss in
Appendix C.

\vskip 1cm
\noindent{\sl Acknowledgements.\quad}
This work is partly supported by Grant-in-Aid for Scientific Research
on Priority Areas 231, the Ministry of Education, Science and Culture.
R.K. and H.Y. are supported by the Japan Society for the Promotion of
Science. O.F. is supported by the Australian Research Council.

\setcounter{equation}{0}

\begin{appendix}
\section{Trigonometric case}
In this appendix we summarize known facts about the $L$-operators and
the vertex operators in the trigonometric case, restricting attention
to the quantum affine algebra $\uq$.
We shall follow the convention of \cite{IIJMNT} concerning $\uq$.
In particular, on the Chevalley generators
$e_i,f_i,t_i=q^{h_i}$ ($i=0,1$) and $q^d$,
the coproduct is chosen to be
\begin{eqnarray*}
&&\Delta(e_i)=e_i\otimes 1+ t_i\otimes e_i,
\\
&&\Delta(f_i)=f_i\otimes t_i^{-1}+ 1\otimes f_i,
\\
&&\Delta(q^h)=q^h\otimes q^h.
\qquad (h=h_0,h_1,d)
\end{eqnarray*}

\subsection{Universal $R$-matrix}
Let $\R$ be the universal $R$-matrix \cite{ICM} for $U=\uq$.
As for its definition and the properties, the reader is referred
e.g. to \cite{ICM,Tanisaki,Nankai}.
For the discussions of $L$-operators one needs to modify
$\R$ slightly \cite{ReshSem}.
Let $\Lambda_i$ denote the fundamental weights ($i=0,1$).
Let further $\rho=\Lambda_0+\Lambda_1$, which we identify with $2d+h_1/2$.
Define
\begin{eqnarray*}
&&
\R^{'+}=q^{(c\otimes\rho+\rho\otimes c)/2}\R,
\\
&&
\R^{'-}=\sigma(\R^{-1})q^{-(c\otimes\rho+\rho\otimes c)/2},
\\
&&\R^{'\pm}(\z)=(\z^\rho\otimes\id)\R^{'\pm}(\z^{-\rho}\otimes\id).
\end{eqnarray*}
Here $\sigma$ stands for the flip of tensor components
$\sigma(a\otimes b)=b\otimes a$.
Then $\R^{'\pm}(\z)$ are formal power series in $\z^{\pm 1}$
of the form
\begin{eqnarray}
&&\R^{'+}(\z)=q^{-h_1\otimes h_1/2}
\left(1-(q-q^{-1})q^2\z \sum_{i=0,1}t_i^{-1}e_i\otimes t_if_i+O(\z^2)
\right),
\nonumber\\
&&\label{Rlead1}\\
&&\R^{'-}(\z)=
\left(1+(q-q^{-1})q^2\z^{-1} \sum_{i=0,1}t_i^{-1}e_i\otimes t_if_i+O(\z^{-2})
\right)q^{h_1\otimes h_1/2}.
\nonumber\\
&&\label{Rlead2}
\end{eqnarray}

The properties of the universal $R$-matrix can be readily translated
in terms of ${\cal R}^{'\pm}$.
For $x\in U$ write $\Delta(x)=\sum x'_i\otimes x''_i$. Then
\beq
\R^{'+}(\z)~\sum\left(\Ad(\z^\rho)x'_i\right)\otimes x''_i
=\sum \Ad(\z^{\rho}q^{c_2\rho/2})x''_i\otimes \Ad(q^{c_1\rho/2})x'_i
{}~\R^{'+}(\z),
\label{RDel1}
\eeq
\beq
\R^{'-}(\z)~\sum \Ad(\z^{\rho}q^{c_2\rho/2})x'_i \otimes
\Ad(q^{c_1\rho/2})x''_i
=\sum \left(\Ad(\z^{\rho})x''_i\right)\otimes x'_i
{}~\R^{'-}(\z).
\label{RDel2}
\eeq
Here $c_1=c\otimes 1$ and $c_2=1\otimes c$.
Similar notations will be used throughout.

The Yang-Baxter equation takes the form
\begin{equation}\label{YBE'}
\begin{array}{cc}
&
\R^{'\pm}_{12}(\z_1/\z_2)
\R^{'\pm}_{13}(q^{\mp c_2/2}\z_1/\z_3)
\R^{'\pm}_{23}(\z_2/\z_3)
\\
&\qquad =
\R^{'\pm}_{23}(\z_2/\z_3)
\R^{'\pm}_{13}(q^{\pm c_2/2}\z_1/\z_3)
\R^{'\pm}_{12}(\z_1/\z_2),
\\
&
\R^{'+}_{12}(q^{c_3/2}\z_1/\z_2)
\R^{'+}_{13}(\z_1/\z_3)
\R^{'-}_{23}(\z_2/\z_3)
\\
&\qquad=
\R^{'-}_{23}(\z_2/\z_3)
\R^{'+}_{13}(\z_1/\z_3)
\R^{'+}_{12}(q^{-c_3/2}\z_1/\z_2).
\end{array}
\end{equation}
For completeness we give
the transformation properties of $\R^{'\pm}$ under the coproduct $\Delta$,
counit $\vep$ and the antipode $a$.
\begin{eqnarray}
&&\left(\Delta\otimes\id\right)\R^{'+}(\z)
=
\R^{'+}_{13}(q^{-c_2/2}\z)\R^{'+}_{23}(\z),
\label{cop1}\\
&&\left(\id\otimes\Delta\right)\R^{'+}(\z)
=
\R^{'+}_{13}(q^{c_2/2}\z)\R^{'+}_{12}(\z),
\label{cop2}\\
&&\left(\Delta\otimes\id\right)\R^{'-}(\z)
=
\R^{'-}_{13}(\z)\R^{'-}_{23}(q^{-c_1/2}\z),
\label{cop3}\\
&&\left(\id\otimes\Delta\right)\R^{'-}(\z)
=
\R^{'-}_{13}(\z)\R^{'-}_{12}(q^{c_3/2}\z),
\label{cop4}\\
&&
(\vep\otimes\id)\R^{'\pm}(\z)=1=(\id\otimes\vep)\R^{'\pm}(\z),
\label{cou}\\
&&
(a\otimes \id)\R^{'\pm}(\z)=\R^{'\pm}(q^{c_1/2}\z)^{-1},
\label{ant1}\\
&&
(\id\otimes a^{-1})\R^{'\pm}(\z)=\R^{'\pm}(q^{- c_2/2}\z)^{-1}.
\label{ant2}
\end{eqnarray}

\subsection{$L$-operators}
Let now
$\pi_V:U'\rightarrow {\rm End}(V)$ be a finite dimensional representation,
where $U'$ signifies the subalgebra of $U=\uq$ generated by
$e_i,f_i,t_i$ ($i=0,1$).
The evaluation representation $\pi_{V_\z}$ associated with $V$
is defined by
\[
\pi_{V_\z}(x)=\pi\left(\z^\rho x \z^{-\rho}\right)
\qquad \forall x\in U'.
\]
Introduce the $L$-operators
\begin{eqnarray*}
&&L^{\pm}(\z)=L_{V}^\pm(\z)=
\left(\pi_{V_\z}\otimes\id\right)\R^{'\pm}.
\\
\end{eqnarray*}

Taking the image of (\ref{YBE'}) in $V_{\z_1}\otimes V_{\z_2}$
we find the following $RLL$ relations:
\begin{eqnarray*}
&&
R^{\pm}_{12}(\z_1/\z_2)\Lpm{1}(\z_1)\Lpm{2}(\z_2)
=
\Lpm{2}(\z_2)\Lpm{1}(\z_1)R^\pm_{12}(\z_1/\z_2),
\\
&&R^{+}_{12}(q^{c/2}\z_1/\z_2)\Lp{1}(\z_1)\Lm{2}(\z_2)
=
\Lm{2}(\z_2)\Lp{1}(\z_1)R^{+}_{12}(q^{-c/2}\z_1/\z_2).
\end{eqnarray*}
Introducing the matrix units $E_{ij}$ let us define the entries
$L_{ij}(\z)$ by
\[
L^\pm(\z)=\sum E_{ij}\otimes L^\pm_{ij}(\z).
\]
In these terms the Hopf algebra structure reads as follows.
\begin{eqnarray*}
&&\Delta \left(L^+_{ij}(\z)\right)
=\sum_l L^+_{lj}(\z)\otimes L^+_{il}(q^{c_1/2}\z),
\\
&&
\Delta \left(L^-_{ij}(\z)\right)
=\sum_l L^-_{lj}(q^{c_2/2}\z)\otimes L^-_{il}(\z),
\\
&&
\vep\left( L^\pm_{ij}(\z)\right)=\delta_{ij},
\\
&&a\left( L^\pm(\z)^t\right)=\left(L^\pm(q^{-c/2}\z)^t\right)^{-1},
\\
&&
a^{-1}\left(L^\pm(\z)\right)={L^\pm(\z)}^{-1}.
\end{eqnarray*}
In the last two lines we set
\begin{eqnarray*}
a\left( L^\pm(\z)^t\right)
&=&\sum E_{ji}\otimes a\left(L^\pm_{ij}(\z)\right),
\\
a^{-1}\left(L^\pm(\z)\right)
&=&\sum E_{ij}\otimes a^{-1}\left(L^\pm_{ij}(\z)\right).
\end{eqnarray*}
All these can be derived by taking the image of
(\ref{cop1})-(\ref{ant2}).

The image of $L^\pm_V(\z_1)$ on the evaluation module $W_{\z_2}$ gives rise
to the $R$-matrix
\begin{eqnarray*}
&&R^{\pm}_{VW}(\z_1/\z_2)
=
\left(\id\otimes\pi_{W_{\z_2}}\right)L^\pm_{V}(\z_1)
=
\left(\pi_{V_{\z_1}}\otimes\pi_{W_{\z_2}}\right)\R^{'\pm}.
\end{eqnarray*}

{}From now on,
$V_\z$ will be the evaluation module associated with
the standard two-dimensional (spin $1/2$) module,
$V=\C v_+\oplus \C v_-$.
In this case
we see from (\ref{Rlead1})-(\ref{Rlead2}) that the $L$-operators have the form
\begin{eqnarray}
L^+(\z)
&=&
q^{c\over4}\left(\matrix{
t_1^{-1/2}+\cdots & \z\, (1-q^2)t_1^{1/2}f_1+\cdots \cr
\z\, (1-q^2)t_0^{1/2}f_0+\cdots & t_0^{-1/2}+\cdots \cr
}\right),
\nonumber\\
L^-(\z)
&=&
q^{-{c\over4}}\left(\matrix{
t_1^{1/2}+\cdots & \z^{-1}\, (1-q^{-2})e_0t_0^{-1/2}+\cdots \cr
\z^{-1}\, (1-q^{-2})e_1t_1^{-1/2}+\cdots & t_0^{1/2}+\cdots \cr
}\right).
\nonumber\\
&&\label{NorL}
\end{eqnarray}

For complex numbers $k,l$, we let $\M_{k,l}^0$ denote
the Verma module over $U$ with highest weight
$(k-l)\Lambda_0 + l \Lambda_1$ and highest weight vector
$\ket{k,l}$ (the superscript $0$ referring to the trigonometric case).
We shall be concerned with the case where $k,l$ are generic, so that
$\M_{k,l}^0$ is irreducible.
{}From (\ref{NorL}) we see in particular that the $0$-th components of the
$L$-operators act on the highest weight vector as
\begin{eqnarray*}
L^+_{\pm\pm,0}\ket{k,l}&=&q^{\pm(k-2l)/4}\ket{k,l},
\\
L^-_{\pm\pm,0}\ket{k,l}&=&q^{\mp(k-2l)/4}\ket{k,l}.
\end{eqnarray*}

Consider the operators
\[
PL^\pm(\z):
v_{\vep'}\otimes u ~ \mapsto ~ \sum_\vep L^\pm_{\vep\vep'}(\z)u \otimes v_\vep
\qquad (u\in \M_{k,l}^0,~v_\vep\in V)
\]
where $P v\otimes u=u\otimes v$.
Taking the image of (\ref{RDel1})-(\ref{RDel2}),
we find that they can be regarded as intertwiners of $U$-modules of the form
\begin{eqnarray*}
PL^+(\z)&:&
V_\z\otimes \M_{k,l}^0 \longrightarrow \M_{k,l}^0\otimes V_{q^{k/2}\z},
\\
PL^-(\z)&:&
V_{q^{k/2}\z}\otimes \M_{k,l}^0 \longrightarrow \M_{k,l}^0\otimes V_{\z}.
\end{eqnarray*}

\subsection{Vertex operators}
Let us proceed to the vertex operators.
They are the intertwiners of $U$-modules of the form
\begin{eqnarray*}
\Phi^{(l',l)}(\z)
&:& \M_{k,l}^0 \longrightarrow \M_{k,l'}^0\otimes V_\z,
\\
\Psi^{*(l',l)}(\z)
&:& V_\z\otimes \M_{k,l}^0 \longrightarrow \M_{k,l'}^0.
\end{eqnarray*}
We call $\Phi$ and $\Psi^*$ vertex operators of type-I and type-II,
respectively.
Nontrivial vertex operators exist if and only if
$l'=l\pm1$,
in which case they are unique up to scalar multiple \cite{FreResh}.
Define their components by
\begin{eqnarray*}
\Phi^{(l',l)}(\z)
&=&\sum\Phi_\vep^{(l',l)}(\z)\otimes v_\vep,
\\
\Psi^{*(l',l)}_\vep(\z)
&=&\Psi^{*(l',l)}(\z)\left(v_\vep\otimes\cdot\right).
\end{eqnarray*}
We have the parity relation
\begin{eqnarray*}
\Phi^{(l\pm 1,l)}_{\vep}(-\z)&=&\mp \vep \Phi^{(l\pm 1,l)}_{\vep}(\z),
\\
\Psi^{*(l\pm 1,l)}_{\vep}(-\z)&=&\pm \vep \Psi^{*(l\pm 1,l)}_{\vep}(\z).
\end{eqnarray*}
We normalize the vertex operators by the conditions
\begin{equation}\label{NorVO}
\begin{array}{ccc}
\Phi^{(l\pm 1,l)}_\mp(\z)\ket{k,l}&=&\ket{k,l\pm 1}+O(\z),
\\
\Psi^{*(l\pm 1,l)}_\pm(\z)\ket{k,l}&=&\ket{k,l\pm 1}+O(\z).
\end{array}
\end{equation}

In terms of the $L^\pm$-operators, the intertwining properties for
the vertex operators read as follows.

\begin{eqnarray*}
\Phim{2}(\z_2)\Lp{1}(\z_1)
&=&
R^+(q^{k/2}\z_1/\z_2)\Lp{1}(\z_1)\Phim{2}(\z_2),
\\
\Phim{2}(\z_2)\Lm{1}(\z_1)
&=&
R^-(\z_1/\z_2)\Lm{1}(\z_1)\Phim{2}(\z_2),
\\
\Lp{2}(\z_2)\Psim{1}(\z_1)
&=&
\Psim{1}(\z_1)\Lp{2}(\z_2)R^{*+}(\z_2/\z_1),
\\
\Lm{2}(\z_2)\Psim{1}(\z_1)
&=&
\Psim{1}(\z_1)\Lm{2}(\z_2)R^{*-}(q^{k/2}\z_2/\z_1).
\end{eqnarray*}
The intertwining relations in this form were presented in  \cite{FreResh}
where the homogeneous grading is used.
Here we have written them in the principal picture.

It is possible to express the type-II vertex operators in terms of type-I
operators and the $L$-operators.
Consider the diagram
\[
\begin{array}{ccc}
   V_\z \otimes \M_{k,l}^0 & \stackrel{\id\otimes\Phi^{(l',l)}(-q^{k/2+1}\z)}
                                   {\longrightarrow}
                     & V_\z \otimes \M_{k,l'}^0 \otimes V_{-q^{k/2+1}\z} \\
                         &        &           \\
    \downarrow \Psi^{*(l',l)}(\z) & & \downarrow PL^+(\z)\otimes\id  \\
                         &        &           \\
\M_{k,l'}^0\otimes \C & \stackrel{\id\otimes{\langle~,~\rangle}}
{\longleftarrow}
     & \M_{k,l'}^0\otimes V_{q^{k/2}\z} \otimes V_{-q^{k/2+1}\z}
\end{array}
\]
%$$\diagram
%{V_\z \otimes \M_{k,l}^0}\ddto^{\Psi^{*(l',l)}(\z)}
%\rrto^{\id\otimes\Phi^{(l',l)}(-q^{k/2+1}\z)}
%&\qquad \qquad \qquad & {V_\z \otimes\M_{k,l'}^0 \otimes V_{-q^{k/2+1}\z}}
%\ddto^{PL^+(\z)\otimes\id} \\
%& &  \\
%\M_{k,l'}^0\otimes \C & &
%\M_{k,l'}^0\otimes V_{q^{k/2}\z} \otimes V_{-q^{k/2+1}\z}
%\llto^{\id\otimes{\langle~,~\rangle}}
%\enddiagram$$
Here the horizontal arrow in the bottom is given by
\[
\langle~,~\rangle:
V_{\z}\otimes V_{-q\z} \longrightarrow \C,
\qquad
v_\vep\otimes v_{\vep'}\mapsto \delta_{\vep,-\vep'}.
\]
All the maps in the diagram are intertwiners.
{}From the uniqueness of the vertex operators we conclude that
\begin{equation}
\Psi^{*(l,l\pm 1)}_{\vep'}(\z)
=q^{\mp(2l-k)/4}
\sum_{\vep} L^+_{\vep\vep'}(\z)\Phi^{(l,l\pm 1)}_{-\vep}(-q^{k/2+1}\z).
\label{Psi=LPhi}
\end{equation}
We have used the normalization (\ref{NorVO}) and (\ref{NorL}) to fix the
scalar multiple in the right hand side.
Notice that the composition in the right hand side
is well defined, because for any vector $\bra{u}$ of the dual
space $\M_{k,l}^{0*}$
the series $\bra{u}L^+_{\vep\vep'}(\z)$ contains
only a finite number of nonzero terms.
In a similar way we find
\begin{equation}
\Psi^{*(l\pm 1,l)}_{\vep'}(\z)
=q^{\mp(2l-k)/4}
\sum_{\vep}\Phi^{(l\pm 1,l)}_{-\vep}(-q^{-k/2-1}\z)
L^-_{\vep\vep'}(q^{-k/2}\z).
\label{Psi=PhiL}
\end{equation}

\subsection{Commutation relations}

The vertex operators are known to
satisfy the commutation relations of the following form \cite{FreResh,IIJMNT}.
\begin{eqnarray}
&&R_{12}(\zeta_1/\zeta_2)
{\Phim{1}}^{(l'',l')}(\zeta_1)
{\Phim{2}}^{(l',l)}(\zeta_2)
\nonumber\\
&&\qquad=
 \sum_{\sigma}
w\left(\matrix{\phantom{l''}l'\phantom{l}\cr
                   l''\phantom{l+\sigma}l \cr
               \phantom{l''}l+\sigma\phantom{l}\cr}
\Biggl|\zeta_1/\zeta_2\right)
{\Phim{2}}^{(l'',l+\sigma)}(\zeta_2)
{\Phim{1}}^{(l+\sigma,l)}(\zeta_1),
\label{triPhiPhi}\\
&&{\Phim{1}}^{(l'',l')}(\zeta_1)
{\Psim{2}}^{(l',l)}(\zeta_2)
\nonumber\\
&&=\qquad
\sum_{\sigma}
\widetilde{w}\left(\matrix{\phantom{l''}l'\phantom{l}\cr
                   l''\phantom{l+\sigma}l \cr
               \phantom{l''}l+\sigma\phantom{l}\cr}
\Biggl|\zeta_1/\zeta_2\right)
{\Psim{2}}^{(l'',l+\sigma)}(\zeta_2)
{\Phim{1}}^{(l+\sigma,l)}(\zeta_1),
\label{triPhiPsi}\\
&&-{\Psim{1}}^{(l'',l')}(\zeta_1)
 {\Psim{2}}^{(l',l)}(\zeta_2)
  \left(R_{12}(\zeta_1/\zeta_2)\right)^{-1}
\nonumber\\
&&\qquad=
\sum_{\sigma}
w\left(\matrix{\phantom{l''}l'\phantom{l}\cr
            l''\phantom{l+\sigma}l \cr
            \phantom{l''}l+\sigma\phantom{l}\cr}
\Biggl|\zeta_1/\zeta_2\right)
\Psim{2}^{(l'',l+\sigma)}(\zeta_2)
\Psim{1}^{(l+\sigma,l)}(\zeta_1).
\label{triPsiPsi}
\end{eqnarray}
Here the $R$-matrix is normalized as
\begin{eqnarray}
&&
R(\z)=\tau^+(\z)^{-1}R^+(\z),
\nonumber \\
\end{eqnarray}
with $\tau^+(\z)$ being defined in (\ref{Rpm}).
The coefficients $w$ are given by
\begin{eqnarray}\label{w}
w\left(\matrix{\phantom{l\pm1}l\phantom{l\mp1}\cr
                  l\pm1\phantom{l}l\mp1 \cr
               \phantom{l\pm1}l\phantom{l\mp1}\cr}
\Biggl|\zeta \right)
&=&
\zeta^{-1}\frac{\xi(\zeta^2;s,q)}{\xi(\zeta^{-2};s,q)},
\\
w\left(\matrix{\phantom{l}l+{\vep'}\phantom{l}\cr
                  l\phantom{l+\vep'}l \cr
               \phantom{l}l+{\vep}\phantom{l}\cr}
\Biggl|\zeta \right)
&=&
\frac{\xi(\zeta^2;s,q)}{\xi(\zeta^{-2};s,q)}
w^{(l)}_{\vep\vep'}(\z),
\end{eqnarray}
where $\xi(z;p,q)$ is defined in (\ref{xi}) and
\begin{eqnarray}
&&w^{(l)}_{\pm\pm}(\z)
=
\frac{\Theta_s(q^2)}{\Theta_s(y^\mp_l)}
\frac{\Theta_s(y^\mp_l\zeta^2)}{\Theta_s(q^2\zeta^2)},
\nonumber\\
&&w^{(l)}_{\pm\,\mp}(\z)
=
\frac{(q^2y^\mp_l;s)_\infty(q^{-2}y^\mp_l;s)_\infty}{(y^\mp_l;s)_\infty^2}
\frac{\Theta_s(\zeta^2)}{\Theta_s(q^2\zeta^2)}\times q\zeta^{-1},
\label{wpm}\\
&&
s=q^{2(k+2)},\qquad
y_l^+=q^{2(l+1)}, \quad
y_l^-=s\left(y_l^+\right)^{-1}.
\nonumber
\end{eqnarray}
The $\widetilde{w}$ are related to $w$ by
\begin{eqnarray}
&&\widetilde{w}\left(\matrix{\phantom{l''}l'\phantom{l}\cr
                  l''\phantom{l^*}l \cr
               \phantom{l''}l^*\phantom{l}\cr}
\Biggl|\zeta \right)
=
\frac{\overline{c}_{l',l}}{\overline{c}_{l'',l^*}}
\tau^{+}(q^{k/2}\z^{-1})
w\left(\matrix{\phantom{l''}\l'\phantom{l}\cr
                  l''\phantom{l^*}l \cr
               \phantom{l''}l^*\phantom{l}\cr}
\Biggl|-q^{-k/2-1}\zeta \right),
\label{wtil}\\
&&
\overline{c}_{l,l+1}=q^{-l/2},
\quad \overline{c}_{l,l-1}=q^{-(k-l)/2}.
\nonumber
\end{eqnarray}

In addition, the type-I vertex operators satisfy
the inversion relation \cite{IIJMNT}
\begin{eqnarray}
g_{l',l}\sum_\vep
\Phi^{(l,l')}_{-\vep}(-q^{-1}\zeta)
\Phi^{(l',l)}_{\vep}(\zeta)
&=&{\rm id}.
\label{triinv1}
\end{eqnarray}
Equivalently it can be put in the form
\begin{eqnarray}
\sum_{l'} g_{l,l'}
\Phi^{(l,l')}_{\vep_1}(\zeta)
\Phi^{(l',l)}_{-\vep_2}(-q^{-1}\zeta)
&=&
\delta_{\vep_1,\vep_2}{\rm id},
\label{triinv2}
\end{eqnarray}
where
\begin{equation}
g_{l\pm 1,l}
=
\frac{(y^\pm_l;s)_\infty}{(q^2y^\pm_l;s)_\infty}
\frac{1}{\widetilde{\xi}(s;s,q)}.
\label{gll}
\end{equation}

Using (\ref{Psi=LPhi})-(\ref{Psi=PhiL}) and (\ref{inv2}),
one can express the $L^\pm$-operators in terms of
type-I and type-II vertex operators.
The result is as follows.
\begin{equation}\label{trigLPsiPhi}
\begin{array}{ll}
L^+_{\vep\vep'}(\z)
&=
\sum_{\pm} q^{\pm(2l-k)/4}g_{l,l\pm 1}\Psi^{*(l,l\pm1)}_{\vep'}(\z)
\Phi^{(l\pm1,l)}_\vep(q^{k/2}\z),
\\
L^-_{\vep\vep'}(\z)
&=
\sum_{\pm} q^{\pm(2l-k)/4}g_{l,l\pm 1}\Phi^{(l,l\pm1)}_{\vep}(\z)
\Psi^{*(l\pm1,l)}_{\vep'}(q^{k/2}\z).
\end{array}
\end{equation}

\medskip\noindent
{\sl Remark.\quad}
The precise meaning of
the commutation relations (\ref{triPhiPhi})--(\ref{triPsiPsi})
is as follows.
For any $\bra{v}\in \M_{k,l''}^{0*}$ and $\ket{u}\in \M_{k,l}^0$
the matrix element
\begin{equation}\label{RHSinPP}
\bra{v}{\Phim{1}}^{(l'',l')}(\zeta_1)
{\Phim{2}}^{(l',l)}(\zeta_2)\ket{u}
\end{equation}
can be meromorphically continued to the entire complex plane
in $\z=\z_2/\z_1$.
Similarly
\begin{equation}\label{LHSinPP}
\bra{v}{\Phim{1}}^{(l'',l')}(\zeta_2)
{\Phim{2}}^{(l',l)}(\zeta_1)\ket{u}
\end{equation}
is meromorphic in $\z^{-1}$.
Equation (\ref{triPhiPhi}) means that for each $\bra{v},\ket{u}$
the matrix elements of the two sides are equal as meromorphic functions.

In fact, equation (\ref{triPhiPhi}) can also be interpreted
as a formal series relation in the following sense.
First note that $R(\z)$ and $w$ have
no poles on the unit circle $|\z|=1$.
Expanding them into Laurent series in the neighborhood of $|\z|=1$,
we can rewrite (\ref{triPhiPhi}) as an equality
\begin{equation}\label{formalser}
\sum_{n_1,n_2}L_{n_1,n_2}\z_1^{-n_1}\z_2^{-n_2}
=
\sum_{n_1,n_2}R_{n_1,n_2}\z_1^{-n_1}\z_2^{-n_2},
\end{equation}
where $L_{n_1,n_2}$ and $R_{n_1,n_2}$ have the form
\[
L_{n_1,n_2}=\sum_{j\in\Z}c_j \Phi_{\vep_1,n_1+j}\Phi_{\vep_2,n_2-j},
\qquad
R_{n_1,n_2}=\sum_{j\in\Z}c'_j \Phi_{\vep_2,n_2-j}\Phi_{\vep_1,n_1+j}.
\]
A close examination shows that the matrix element
(\ref{RHSinPP})
is holomorphic in the domain $|\z^2|<q^{-4}$,
while
(\ref{LHSinPP})
is holomorphic in $|\z^2|>q^{4}$.
(This analysis requires the use
of the explicit formula for the two-point functions
obtained by solving the $q$-KZ equation.)
Hence the matrix elements of both sides of (\ref{formalser})
converge absolutely to the same quantity in a common domain containing
$|\z|=1$.
It follows that $L_{n_1,n_2}$, $R_{n_1,n_2}$ are well defined operators and
that $L_{n_1,n_2}=R_{n_1,n_2}$ for any $n_1,n_2$.

The formal series relation is thus a stronger statement,
and its validity relies on the analyticity structure of the matrix elements.
It is {\it not} valid e.g. in the case (\ref{triPsiPsi}).

We remark also that in the trigonometric case
the choice of the Riemann-Hilbert factorization
in the $\Phi\Phi$ ordering rule (\ref{phiphi2}) is determined by a similar
principle.
In the right hand side of (\ref{phiphi2}) the pole corresponding to
$1+q\z_1/\z_2$ is spurious since $\beta_1(\z_1/\z_2)$ contains the same
factor.
On the other hand the factor $1+q^{-1}\z_2/\z_1$ is not contained in
$\beta_1(\z_2/\z_1)$.
Nevertheless
one can show that the matrix elements of
$\sum_{m\not \equiv n~(2)}
\varphi^{(l\pm1,l)}_m\varphi^{(l,l\mp1)}_n \zeta_1^{-m}\zeta_2^{-n}$
is divisible by $1+q^{-1}\z_2/\z_1$.
Therefore (\ref{phiphi2}) is correct as a formal series relation.

\setcounter{equation}{0}

\section{Two-point functions}

In order to determine the proportionality constants $a$ of
(\ref{a-const}), $c_{l,l\pm 1}$ of (\ref{c-const}) an $g_{l\pm 1,l}$
of (\ref{g-def}), it is necessary to compute the appropriate two-point
functions. We outline the method of calculation below, and list the
required results.

Consider the two-point function
$\sum_{\vep}\langle
\Phi^{(l\pm 1,l)}_{\vep}(\z_2)
\Phi^{(l,l\mp 1)}_{\vep}(\z_1)\rangle$.
By using the normal-ordering rule (\ref{Phinormal}), we find
\[
\langle \varphi^{(l\pm1,l)}_m\varphi^{(l,l\mp1)}_n \rangle
=\delta_{m,-n}e_{2m,0}
=\delta_{m,-n}\widetilde{\alpha}_{1,m},
\]
which can be summed to give the two-point function,
\[
\sum_{\vep}\langle
\Phi^{(l\pm 1,l)}_{\vep}(\z_2)
\Phi^{(l,l\mp 1)}_{\vep}(\z_1)\rangle
=\alpha_1(\z)^{-1}.
\]

Alternatively the same result can be obtained from the following argument.
Taking the expectation value of the commutation relation
(\ref{phiphi1}) we have
\begin{equation}\label{vacpp}
\alpha_1(\zeta_1/\zeta_2)
\sum_{m\equiv n~(2)}\langle
\varphi^{(l\pm1,l)}_m\varphi^{(l,l\mp1)}_n \zeta_2^{-m}\zeta_1^{-n}
\rangle=
(\zeta_2\leftrightarrow \zeta_1).
\end{equation}
The left hand side (resp. right hand side) of (\ref{vacpp})
is a power series in $\z=\z_1/\z_2$ (resp. $\z^{-1}$).
Hence, assuming they have a common domain of convergence,
we conclude that the left hand side of (\ref{vacpp}) is
a constant. {}From the normalization (\ref{phinormalization}) of
the vertex operators we find this constant to be 1.

Using a similar procedure, one can compute the other two-point
functions, which we list below. In practical
calculations, it is convenient to define a `dagger-rule', a $B$-linear
anti-automorphism defined by
\[
(L^{(l)\e}_n)^\dagger = (-p^{1/2})^n L^{(l)\e}_{-n},\qquad
(\varphi_n^{(l,l')})^\dagger = \varphi_{-n}^{(l',l)}.
\]
The ordering rules are preserved by the action of $\dagger$, and
hence
\[
\langle a \rangle = \langle a^\dagger\rangle \qquad \forall a\in \As.
\]

Let
\begin{eqnarray*}
&&f^{(l)\pm}_{1}(z)=
{}_2\phi_1\left(\matrix
{q^2 \phantom{y^\pm_l} q^2y^\pm_l \cr
\phantom{q^2}y^\pm_l\phantom{q^2y^\pm_l}\cr}
;s,q^{-2}z\right), \\
&&f^{(l)\pm}_{2}(z)=
-\frac{1-q^2}{1-y^\pm_l}
q^{-1}{}_2\phi_1\left(\matrix
{s q^2 \phantom{s y^\pm_l} q^2y^\pm_l \cr
\phantom{s q^2}s y^\pm_l\phantom{q^2y^\pm_l}\cr}
;s,q^{-2}z\right),
\end{eqnarray*}
where the basic hypergeometric function is
\begin{equation}\label{hyper}
{}_2\phi_1\left(\matrix
{a \phantom{c} b \cr
\phantom{a}c\phantom{b}\cr};q,z\right)
=\sum_{n=0}^\infty
\frac{(a;q)_n(b;q)_n}{(c;q)_n(q;q)_n}z^n,
\qquad
(a;q)_n=\prod_{j=0}^{n-1}(1-aq^j)~.
\end{equation}
The functions $f^{(l)\pm}_i(z)$ are related to those of (\ref{f-def})
by
$$
f^{(l)\pm}_1(s\z^2)+y_l^\pm\z f^{(l)\pm}_2(s\z^2)=
\frac{1}{1+q\z}\left(f^{(l)}_{+\pm}(\z)+f^{(l)}_{-\pm}(\z)\right)~.
$$
Below, $\z=\zeta_1/\zeta_2$.
We have
\begin{eqnarray}
&&\sum_{\vep} \langle \Phi^{(l\pm 1,l)}_{\vep}(\z_2)
\Phi^{(l,l\mp 1)}_{\vep}(\z_1)\rangle
=\frac{(p^{1/2}q^{-1}\zeta;p)_\infty}{(p^{1/2}q\zeta;p)_\infty}
\frac{\xi(\zeta^2;p,q)}{\xi(\zeta^2;s,q)}~,\\
&&\sum_{\vep} \langle \Phi^{(l,l\pm 1)}_{-\vep}(\z_2)
\Phi^{(l\pm 1,l)}_{\vep}(\z_1)\rangle
{\begin{array}[t]{l}=\displaystyle
\frac{(-pq^{-1}\zeta;p)_\infty}{(-pq\zeta;p)_\infty}
\frac{\xi(\zeta^2;p,q)}{\xi(\zeta^2;s,q)}\cr\displaystyle\times \left(
f^{(l)\pm}_1(s \zeta^2)+y^\pm_l\zeta\times f^{(l)\pm}_2(s \zeta^2)\right),
\cr\end{array}}\label{Phi2}
\\
&&\sum_{\vep} \langle \Psi^{*(l\pm 1,l)}_{-\vep}(\z_2)
\Phi^{(l,l\mp 1)}_{\vep}(\z_1)\rangle{\begin{array}[t]{l}=\displaystyle
\sum_{\vep} \langle \Phi^{(l\pm 1,l)}_{-\vep}(\z_2)
\Psi^{*(l,l\mp 1)}_{\vep}(\z_1)\rangle\cr
=\displaystyle\frac{\xi(s^{-1/2}\zeta^2;0,q)}{\xi(s^{-1/2}\zeta^2;s,q)},
\cr
\end{array}
}
\\
&&\sum_{\vep} \langle \Psi^{*(l,l\pm 1)}_{\vep}(\z_2)
\Phi^{(l\pm 1,l)}_{\vep}(\z_1)\rangle
=\sum_{\vep} \langle
\Phi^{(l,l\pm 1)}_{\vep}(\z_2)
\Psi^{*(l\pm 1,l)}_{\vep}(\z_1)\rangle
\\
&&\quad=\frac{\xi(s^{-1/2}\zeta^2;0,q)}{\xi(s^{-1/2}\zeta^2;s,q)}
\times \left(
f^{(l)\pm}_1(s^{1/2}\zeta^2)-(y^\pm_l)^{1/2}\zeta\times
f^{(l)\pm}_2(s^{1/2}\zeta^2)
\right),\nonumber
\\
&&\sum_{\vep} \langle \Psi^{*(l\pm 1,l)}_{\vep}(\z_2)
\Psi^{*(l,l\mp 1)}_{\vep}(\z_1)\rangle
{\begin{array}[t]{l}=\displaystyle\frac{(p^{*1/2}q\zeta;p^*)_\infty}
{(p^{*1/2}q^{-1}\zeta;p^*)_\infty}\cr\times\displaystyle
\frac{1-\zeta^2}{\xi(\zeta^2;p^*,q)\xi(\zeta^2;s,q)},\cr\end{array}}\\
&&\sum_{\vep} \langle
\Psi^{*(l,l\pm 1)}_{-\vep}(\z_2)\Psi^{*(l\pm 1,l)}_{\vep}(\z_1)\rangle\\
&&\quad=\frac{(-p^*q^{-1}\zeta;p^*)_\infty}{(-p^*q\zeta;p^*)_\infty}
\frac{1-\zeta^2}{\xi(\zeta^2;p^*,q)\xi(\zeta^2;s,q)}
\left(f^{(l)\pm}_1(\zeta^2)+\zeta\times f^{(l)\pm}_2(\zeta^2)\right),
\nonumber\\
&&\sum_{\vep'} \langle \Phi_{\sigma\vep'}^{(l,l')}(\zeta_2)
L_{\vep'\vep}^{(l')}(\zeta_1) \rangle=
a^{(l)}_\vep\alpha^{(2)\sigma}(\z)^{-1}, \\
&&\sum_{\vep'}\langle L_{\vep'\vep}^{(l)}(\z_2)
\Phi^{(l,l')}_{\sigma\vep'}(\z_1) \rangle=
a^{(l)}_\vep\alpha^{(1)\sigma}(\z)^{-1},\\
&&\sum_{\vep'} \langle L_{\vep\vep'}^{(l)}(\zeta_2)
\Psi_{\sigma\vep'}^{*(l,l')}(\zeta_1) \rangle=
a^{(l)}_\vep\alpha^{*(1)\sigma}(q^{k/2}\z),\\
&&\sum_{\vep'} \langle\Psi^{*(l,l')}_{\sigma\vep'}(\z_2)
L^{(l')}_{\vep\vep'}(\z_1) \rangle=
a^{(l)}_\vep\alpha^{*(2)\sigma}(q^{-k/2}\z),\\
&&
\sum_{\vep,\vep'} \langle L_{\vep\vep'}^{(l)}(\z_2)
L_{\vep\vep'}^{(l)}(\z_1) \rangle=
a^2(q^{k-l}+q^l)\frac{\overline{\alpha}^*(\z)}{\overline{\alpha}(\z)},
\\
&&\sum_{\vep,\vep'} \langle L_{\vep\vep'}^{(l)}(\z_2)
L_{-\vep-\vep'}^{(l)}(\z_1) \rangle=
2a^2q^{k/2}\frac{\overline{\beta}^*(\z)}{\overline{\beta}(\z)}, \label{L2}
\end{eqnarray}
where $l'=l\pm 1$, $l-l'=\sigma \e$ and $a^{(l)}_\e$ is defined in
(\ref{al-def}) below.
The functions
$\xi(\z^2;p,q), \rho(\z^2), \overline{\alpha}(\z), \overline{\beta}(\z),
\alpha^{(1)\sigma}(\z), \alpha^{(2)\sigma}(\z)$,
are defined in (\ref{xi}), (\ref{rho})--(\ref{beta})
and (\ref{ab-func}), respectively.

To determine the constant $g_{l,l'}$ in (\ref{g-def}), we use
(\ref{Phi2}) with $\z_1=\z, \z_2=-q^{-1}\z$ to obtain
\begin{eqnarray*}
\sum_{\vep} \langle
\Phi^{(l,l\pm 1)}_{-\vep}(-q^{-1}\z)
\Phi^{(l\pm 1,l)}_{\vep}(\z)\rangle&& \nonumber\\&&\hbox{\hspace{-1.5in}}=
\frac{(s;s)_\infty}{(sq^2;s)_\infty}
\frac{\widetilde{\xi}(s;s,q)}{\widetilde{\xi}(p;p,q)}
\left( f^{(l)\pm}_1(sq^2)-qy_l^{\pm}f^{(l)\pm}_2(sq^2) \right).
\end{eqnarray*}
The expression in (\ref{g-def}) results from the identity
\[
f^{(l)\pm}_1(sq^2)-qy_l^{\pm}f^{(l)\pm}_2(sq^2)=
\frac{(sq^2;s)_\infty}{(s;s)_\infty}
\frac{(q^2y_l^{\pm};s)_\infty}{(y_l^{\pm};s)_\infty}.
\]
The constants $a$ and $c_{l,l'}$ are be determined
in a similar manner.

Finally let us comment on the normalization (\ref{hwcond}) of the
$L$-operator. Let
\[L^{(l)\pm}_0|k,l\rangle =a_{\pm}^{(l)}|k,l\rangle.\]
{}From the requirement that $\qdet L(\z)=q^{k/2}$, using (\ref{L2}), we have
\beq a^{(l)}_+a^{(l)}_-=a^2q^{k/2}. \label{eig1}\eeq
In addition, taking the expectation value of (\ref{lphi-split}) we
find
$$
\frac{\overline{\alpha}(\z)}{\rho(\z^2)}
\sum_{\vep}\langle \Phi_{\vep}^{(l\pm 1,l)}(\z_2)L^{(l)}_{\vep\e'}(\z_1)\rangle
=\overline{\alpha}(\z^{-1})\sum_{\vep}\langle
L^{(l\pm 1)}_{\vep\e'}(\z_1)\Phi^{(l\pm 1,l)}_{\vep}(\z_2)\rangle.
$$
Again the left hand side is a power series in $\z$ while the right hand
is a power series in $\z^{-1}$, with the expansions
\[
\mbox{LHS}= q^{1/2}a^{(l)}_{\mp}+O(\z),\qquad
\mbox{RHS}=a^{(l\pm 1)}_{\mp}+O(\z^{-1}).
\]
We conclude that both sides are constant, and therefore
\beq\label{eig2}
q^{1/2}a^{(l)}_{\mp}=a^{(l\pm 1)}_{\mp}.
\eeq
The two conditions
(\ref{eig1}) and (\ref{eig2}) motivate our choice of normalization
\beq\label{al-def}
a_+^{(l)}=aq^{(k-l)/2}, ~~a_-^{(l)}=aq^{l/2}~.
\eeq

\section{The Shapovalov-type determinant}
\setcounter{equation}{0}

Using the ordering rules derived in Sections 3, 5 and 6, it is
possible to compute a Shapovalov determinant for the operators
introduced above, for low degree $d$. This is the determinant of a
matrix formed by coupling of vectors in $\MKL$ with those in $\DMKL$.
It appears that for all three of the conjectural basis vectors
presented above, (\ref{LBASIS}), (\ref{basis}), or (\ref{basis1}) in
the level-one case, this determinant has a simple factorized form to
the degree we computed it.

Consider first the $N_d\times N_d$ matrix (with $N_d$ as in (\ref{Nd}))
formed by coupling degree-$d$ vectors (\ref{LBASIS}) in $\MKL$,
generated by Laurent components of the $L$ operator, with vectors
(\ref{DLbasis}).  The elements of this matrix are complicated
functions of the four variables $q,p,l,k$. However, The determinant
$\Delta_d$ of this matrix has the following simple factorized form for
degrees 1 and 2:
\begin{eqnarray}
\Delta_1& =&\frac{a^4 p
(q-q^{-1})^4}{(p;p)_1(p^*;p^*)_1}\, [k-l][l],\nonumber\\
\Delta_2&=&\frac{a^{12}q^{-k}p^4
(q-q^{-1})^{14}}
{(p;p)_1(p;p)_2(p^*;p^*)_1(p^*;p^*)_2}\, [2][k-l-1][l-1][k-l]^2[l]^2[k+2],
\nonumber\\
\label{ldet}
\end{eqnarray}
where $$ [n]=\frac{q^n-q^{-n}}{q-q^{-1}} $$ and $a$ is defined in
(\ref{a-const}). Note that the determinant factorizes into a factor of
the form $f(p)f(p^*)$ and a factor independent of $p$. The dependence
on $q$ is essentially in the form of the Shapovalov
determinant~\cite{KacKaz,DeCKac}.  In the trigonometric case, the
zeros of the Shapovalov determinant correspond to special values of
$(k,l)$ where the Verma module has null vectors. The above results
suggest that the structure of the Verma module is unchanged in this
sense under $p$-deformation.

The determinants of matrices formed by coupling degree-$d$
$\varphi$-vectors (\ref{basis}) in $\MKL$ with (\ref{dualbasis}) in
$\DMKL$ also have a simple factorized form:
\begin{eqnarray}
\Delta_1
&=&(1)\,
\frac{1}{[l][l-k]},\nonumber\\
\Delta_2
&=& (1)(2)\,
\frac{[2]}{[k+2][l]^2[l-1][k-l]^2[k-l-1]},\label{phidet}
\end{eqnarray}
where
\[
(n)=\frac{(p^*;p^*)_n}{(p;p)_n}\, q^{\frac{n(n+1)k}{2}}.
\]
The factor in the determinants (\ref{phidet}) containing the variable
$q$ is the inverse of the Shapovalov determinant.
We expect such a structure persists for higher degrees.

For level-one, the determinant of the matrix of the coupling between
${\cal H}^{(i)}_d$ and its dual ${\cal H}^{(i)*}_d$ with respect to
the basis (\ref{basis1}) and the corresponding one for ${\cal
H}^{(i)*}_d$ small $d$ also has a simple form:
\begin{eqnarray*}
\begin{array}{ll}
\Delta_1&=(1)
\\
\Delta_2&=(2)\,\displaystyle\frac{1}{[2]}
\\
\Delta_3&=(1)(3)\,\displaystyle\frac{1}{[2]}
\\
\Delta_4&=(1)^2(4)\,\displaystyle\frac{1}{[2][4]}
\\
\Delta_5&=(1)^2(2)(5)\,\displaystyle\frac{1}{[2]^2[4]}
\\
\Delta_6&
=(1)^3(2)(3)(6)\, \displaystyle\frac{[3]}{[2]^3[4][6]}
\\
\Delta_7&
=(1)^4(2)^2(4)(7)\,\displaystyle\frac{[3]}{[2]^4[4]^2[6]}
\\
\Delta_8&
=(1)^6(2)^2(3)(5)(8)\, \displaystyle\frac{[3]}{[2]^5[4]^2[6][8]}
\end{array}
\end{eqnarray*}

We remark that such a factorization into a function of $q$ and a
function of $p$ is reminiscent of that for the spontaneous staggered
polarization \cite{BaxKel,8v}.
\end{appendix}
\bibliographystyle{unsrt}

\end{document}